\documentclass[aps,prc,twocolumn,superscriptaddress,showpacs,showkeys]{revtex4}
\usepackage{amsmath,amssymb}
\usepackage{bm}
\usepackage[dvips]{graphicx}
\input{epsf}
\newcommand{\re}{\text{Re}}

\newcommand{\Tr}{\text{Tr}}
\newcommand{\vectau}{\vec\tau}
\newcommand{\vecpi}{\vec\pi}
\newcommand{\del}{\partial}
\newcommand{\Nmass}{m_{N}}
\newcommand{\NstarMass}{m_{N^{*}}}
\newcommand{\NstarWidth}{\Gamma_{N^{*}}}
\newcommand{\be}{\begin{eqnarray}}
\newcommand{\ee}{\end{eqnarray}}

\newcommand{\ket}{\rangle}
\newcommand{\bra}{\langle}

\newcommand{\ProtonMass}{M_{P}}

\begin{document}

\title{Magnetic moments of the \( N(1535) \) resonance
in the chiral unitary model}

\author{T.~Hyodo}%
\affiliation{Research Center for Nuclear Physics (RCNP), 
Ibaraki, Osaka 567-0047, Japan}%
\author{S.~I.~Nam}%
\affiliation{Research Center for Nuclear Physics (RCNP), 
Ibaraki, Osaka 567-0047, Japan}%
\affiliation{Department of Physics, 
Pusan National University, Pusan 609-735, Korea}%
\author{D.~Jido}%
\altaffiliation[Present address: ]{
ECT*, European Centre for Theoretical Studies
in Nuclear Physics and Related Areas
Villa Tambosi, Strada delle Tabarelle 286,
I-38050 Villazzano (Trento), Italy}
\affiliation{Research Center for Nuclear Physics (RCNP), 
Ibaraki, Osaka 567-0047, Japan}%
\author{A.~Hosaka}%
\affiliation{Research Center for Nuclear Physics (RCNP), 
Ibaraki, Osaka 567-0047, Japan}%

\date{\today}% It is always \today, today,
             %  but any date may be explicitly specified

\begin{abstract}
    We calculate the magnetic moments of the \( N(1535) \)
    resonance using the chiral unitary model,
    where the resonance is dynamically generated
    in the scatterings
    of the lowest-lying mesons and baryons.
    We obtain the magnetic moments of the resonance 
    as $+1.1$ and $-0.25$
    for $p(1535)$ and $n(1535)$, respectively,
    in units of the nuclear magneton.
    We discuss the origin of these numbers
    within the chiral unitary model,
    then we compare the present results 
    with those of the quark model and the chiral doublet model.
    The possibility to observe the magnetic moments in experiments
    is also investigated.
\end{abstract}


\maketitle

%%%%%%%%%%%%%%%%%%%%%%%%%%%%%%%%%%%%%%%%%%%%%%%%%%%%%%%%
\section{Introduction}\label{sec:intro}

% physical background of study of N(1535)
The study of the properties of baryon resonances has attracted continuous
attention and is one of the most important topics in hadron physics. 
The first nucleon resonance with the negative parity, $N(1535)$, has
unique feature of its strong coupling to the $\eta N$ state, which allows
us to take relatively 
clean data 
%in experiments
to other resonance
regarding the eta meson in the final state as a probe of the intermediate 
$N(1535)$. Theoretically hadronic resonances have been investigated in recent
lattice calculations~\cite{Nakahara:1999vy,Sasaki:2001nf,Sasaki:2002sj},
and direct comparison with QCD is becoming possible. 
On the other line, the property of the negative parity resonance could be related to
chiral symmetry of QCD, in particular, to symmetry restoration
at finite temperature and density, as reported in the linear representations
of chiral symmetry for the resonance~\cite{DeTar:1989kn,Jido:1998av,Jido:2001nt}.

% chiral unitary model
Recently,  
the chiral unitary model has been successfully applied to meson-baryon
scatterings and to the description of the $s$-wave baryon
resonances as dynamically generated objects by
the ground state mesons and baryons~\cite{Kaiser:1995cy,
Kaiser:1995eg,Kaiser:1997js,Krippa:1998us,Oset:1998it,Lutz:2001yb}.
For instance, $\Lambda(1405)$ being treated as a $\bar K N$
and $\pi \Sigma$ state reproduces well experimental data available now.
Further, $N(1535)$ and other states in low lying $J^{P}=1/2^{-}$
states are also discussed in the same
framework~\cite{Inoue:2001ip,Oset:2001cn,Ramos:2002xh}.

% magnetic moments
In order to investigate the electromagnetic structure of the dynamically
generated resonances, in Ref.~\cite{Jido:2002yz},
the magnetic moments were calculated for $\Lambda(1405)$ and $\Lambda(1670)$
in the chiral unitary model.
Following that paper,
we will here calculate the magnetic moments of another
$J^{P}=1/2^{-}$ resonance state, $N(1535)$,
in the chiral unitary model.
Recently, the resonance magnetic moments were studied also
in the quark model~\cite{Chiang:2002ah},
and therefore, the present study
provides one of the alternative descriptions.

% experiments
Experimentally, magnetic moments of resonance states 
can be extracted through bremsstrahlung processes.
So far, the magnetic moments of $\Delta^{++}(1232)$
have been studied in the reaction
$\pi^{+}p\to \gamma \pi^{+}p$~\cite{Nefkens:1978eb,Bosshard:1991zp},
where $\mu_{\Delta^{++}}= 3.7\sim 7.5\mu_{N}$ was
extracted~\cite{Hagiwara:2002fs}.
The uncertainty in the number arises from the ambiguity
in the theoretical analysis of the reaction.
Now for $N(1535)$, a similar process can be used such as
$\gamma p \to \gamma \eta p$~\cite{Chiang:2002ah}.

% organization of the paper
This paper is organized as follows.
In section~\ref{sec:formulation},
We present the formulation to calculate the magnetic moments
in the chiral unitary model.
The input parameters and numerical results are presented
in section~\ref{sec:Nresults}.
In section~\ref{sec:discussion}
we discuss the obtained results from various point of view.
In section~\ref{sec:exp}, we discuss the possibility to observe the
magnetic moments of $N^{*}$ in experiment and calculate the
energy spectra and the angular distributions of the emitted photon 
in the $\gamma N \rightarrow \gamma \eta N$ and 
$\pi^- p \rightarrow \gamma \eta n$ reactions. 
Section~\ref{sec:summary} is devoted as the summary
of the present results.

%%%%%%%%%%%%%%%%%%%%%%%%%%%%%%%%%%%%%%%%%%%%%%%%%%%%%%%%
\section{Formulation}\label{sec:formulation}

\begin{figure}[tbp]
    \centering
    \includegraphics[width=8.3cm,clip]{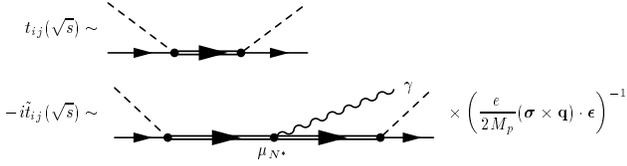}
    \caption{Feynman diagrams of the amplitudes
    $t_{ij}(\sqrt{s})$ and $-i\tilde{t}_{ij}(\sqrt{s})$
    around energy region of the resonance.
    Solid, dashed, wavy and double lines represent
    baryons, mesons, photon and baryon resonances, respectively.
    In calculating $-i\tilde{t}_{ij}(\sqrt{s})$,
    we consider the diagrams which
    contribute to the magnetic moments,
    and extract a factor
    in order to make the coupling of resonance
    to photon to be magnetic moment
    in units of the nuclear magneton.}
    \label{fig:MMamplitudes}
\end{figure}

In this section, 
we briefly explain the chiral unitary model,
where we obtain the amplitude of meson-baryon scatterings.
The amplitude of $\pi N$ scatterings
obtained in the chiral unitary model
can be interpreted as the resonance pole term
around the $N(1535)$ resonance energy region,
as shown in the upper diagram in Fig.~\ref{fig:MMamplitudes},
We then introduce the photon field and
electromagnetic couplings
based on effective chiral Lagrangian,
and we calculate the amplitude
for the photon-resonance coupling
as shown in Fig.~\ref{fig:MMamplitudes}.
Using these two amplitudes, we evaluate
the magnetic moments of the resonance
in two different ways.
Here we follow the formulation
in Ref.~\cite{Jido:2002yz}.

\subsection{Chiral unitary model}\label{subsec:ChU}

Chiral unitary model is an extension of 
the chiral perturbation theory to the
resonance energy region, by imposing the unitarity condition.
%In this work, we adopt 
Using
the N/D method~\cite{Oller:2000fj,Meissner:1999vr}
for the unitarity condition,
the T-matrix amplitude can be written as
\begin{align}
    T=[1-VG]^{-1}V \ ,
    \label{eq:Tmat}    
\end{align}
with the basic interaction $V$ 
and the loop function $G$,
which are given in the following.
This equation provides algebraically
the solution to the Bethe-Salpeter equation.

We derive the basic interaction $V$ from the
chiral perturbation theory,
which well describes low-energy hadron dynamics.
For the $J^P=1/2^-$ baryons, 
the $s$ wave scatterings are relevant,
for which the basic meson-baryon interactions
are given by the Weinberg-Tomozawa term.
The non-relativistic form of the interaction term
is then given by
\begin{align}
    V_{ij}
    =&-\frac{C_{ij}}{4f_if_j}
    (2\sqrt{s}-M_i-M_j)\sqrt{\frac{E_i+M_i}{2M_i}}
    \sqrt{\frac{E_j+M_j}{2M_j}} \ ,
    \label{eq:WTint} 
\end{align}
where the indices $(i,j)$ denote the channels
of meson-baryon scatterings
and the coefficients $C_{ij}$ are fixed by chiral symmetry.
In this work we are interested in the \( N(1535) \) resonance,
and therefore, we calculate the scattering amplitude
with the strangeness \( S=0 \)
and the electric charge $Q=0$ and $Q=1$.
The relevant channels and the coefficients $C_{ij}$
are given in Table~\ref{tbl:Cij}.
\begin{table*}
    \caption{$C_{ij}(S=0,Q=0)$ and $C_{ij}(S=0,Q=1)$}
    \label{tbl:Cij}
\begin{ruledtabular}
\begin{tabular}{ccccccc|ccccccc}
    \multicolumn{7}{c|}{$C_{ij}(S=0,Q=0)$}
    & \multicolumn{7}{c}{$C_{ij}(S=0,Q=1)$}  \\ 
    & $\pi^{0}n$ & $\pi^{-}p$  & $\eta n$
    & $K^{0}\Lambda$ & $K^{0}\Sigma^{0}$ & $K^{+}\Sigma^{-}$
    & & $\pi^{0}p$ & $\pi^{+}n$  & $\eta p$
    & $K^{+}\Lambda$ & $K^{+}\Sigma^{0}$ & $K^{0}\Sigma^{+}$ \\ \hline
%%%%%%%%%%%%%%%%%%%%%%%%%%%%%%%%%%%%%%%%%%%%%%%%%%%%%%%%%%%%%%%%%%%%%
$\pi^{0}n$ & $0$ & $-\sqrt{2}$ & $0$
     & $\frac{\sqrt{3}}{2}$ & $-\frac{1}{2}$
     & $-\frac{1}{\sqrt{2}}$
& $\pi^{0}p$ & $0$ & $\sqrt{2}$ & $0$
     & $-\frac{\sqrt{3}}{2}$ & $-\frac{1}{2}$
     & $\frac{1}{\sqrt{2}}$ \\
%%%%%%%%%%%%%%%%%%%%%%%%%%%%%%%%%%%%%%%%%%%%%%%%%%%%%%%%%%%%%%%%%%%%%
$\pi^{-}p$   & & $1$ & $0$ & $-\sqrt{\frac{3}{2}}$
     & $-\frac{1}{\sqrt{2}}$ & $0$
& $\pi^{+}n$ & & $1$ & $0$ & $-\sqrt{\frac{3}{2}}$
     & $\frac{1}{\sqrt{2}}$ & $0$ \\
%%%%%%%%%%%%%%%%%%%%%%%%%%%%%%%%%%%%%%%%%%%%%%%%%%%%%%%%%%%%%%%%%%%%%
$\eta n$   & & & $0$ & $-\frac{3}{2} $
       & $-\frac{\sqrt{3}}{2}$ & $-\sqrt{\frac{3}{2}}$
& $\eta p$ & & & $0$ & $-\frac{3}{2} $
       & $\frac{\sqrt{3}}{2}$ & $-\sqrt{\frac{3}{2}}$ \\
%%%%%%%%%%%%%%%%%%%%%%%%%%%%%%%%%%%%%%%%%%%%%%%%%%%%%%%%%%%%%%%%%%%%%
$K^{0}\Lambda$ & & & & $0$ & $0$ & $0$
& $K^{+}\Lambda$ & & & & $0$ & $0$ & $0$ \\
%%%%%%%%%%%%%%%%%%%%%%%%%%%%%%%%%%%%%%%%%%%%%%%%%%%%%%%%%%%%%%%%%%%%%
$K^{0}\Sigma^{0}$ & & & & & $0$ & $-\sqrt{2}$
& $K^{+}\Sigma^{0}$ & & & & & $0$ & $\sqrt{2}$ \\
%%%%%%%%%%%%%%%%%%%%%%%%%%%%%%%%%%%%%%%%%%%%%%%%%%%%%%%%%%%%%%%%%%%%%
$K^{+}\Sigma^{-}$ & & & & & & $1$
& $K^{0}\Sigma^{+}$ & & & & & & $1$\\
\end{tabular}
\end{ruledtabular}
\end{table*}
In Eq.~\eqref{eq:WTint},
\( f_i \), $M_i$, and $E_i$ 
are the meson decay constant,
the observed baryon mass, and the energy
of the baryon in the channel $i$, respectively.
\( s \) denotes the total energy in the center of mass system.

In order to calculate the loop-integral function $G$,
we employ the dimensional regularization:
\begin{equation}
    \begin{split}
    G_i(\sqrt{s})
    =&i\int\frac{d^{4}q}{(2\pi)^{4}}
    \frac{2M_i}{(P-q)^{2}-M_i^{2}}
    \frac{1}{q^{2}-m_i^{2}} \\
    =&\frac{2M_{i}}{(4\pi)^{2}}
    \Biggl\{a_i(\mu)+\ln\frac{M_i^{2}}{\mu^{2}}
    +\frac{m_i^{2}-M_i^{2}+s}{2s}\ln\frac{m_i^{2}}{M^{2}_i} \\
    &
    +\frac{\bar{q}_i}{\sqrt{s}}
    \Bigl[\ln(s-(M^{2}_i-m_i^{2})+2\sqrt{s}\bar{q}_i) \\
    &
    +\ln(s+(M^{2}_i-m_i^{2})+2\sqrt{s}\bar{q}_i) \\
    &
    -\ln(-s+(M^{2}_i-m_i^{2})+2\sqrt{s}\bar{q}_i) \\
    &
    -\ln(-s-(M^{2}_i-m_i^{2})+2\sqrt{s}\bar{q}_i)
    \Bigr]\Biggl\},
    \end{split}
    \label{eq:loop}
\end{equation}
where \( m_i \) is the mass of the meson in channel \( i \).
The three-momentum of the intermediate meson
\( \bar{q}_i \) is defined by
\begin{equation}
    \bar{q}_{i}(\sqrt{s})=\frac{\sqrt{(s-(M_i-m_i)^{2})
    (s-(M_i+m_i)^{2})}}{2\sqrt{s}} \ .
    \label{eq:qdef}
\end{equation}
In Eq.~\eqref{eq:loop},
$\mu$ and \( a_i(\mu) \) 
are the regularization scale and
the subtraction constants,
which will be taken as free parameters of this model.

Substituting Eqs.~\eqref{eq:WTint} and \eqref{eq:loop}
into Eq.~\eqref{eq:Tmat},
we obtain the T-matrix amplitude of meson-baryon scatterings $t_{ij}$.
The advantage of this model is that
we obtain the amplitude $t_{ij}$ in analytic form,
so that it can be extended to the complex energy plane.
When the subtraction constants \( a_i \) are properly fixed,
the amplitude $t_{ij}$ provides 
a good description for observables such as
cross sections and phase shifts~\cite{Inoue:2001ip,Oset:2001cn,Ramos:2002xh}.

In the present model the pole of the amplitude can be calculated 
in the complex plane.
From this pole, we extract the information of the resonance.
Around the resonance region $\sqrt{s}\sim M_{N^*}$,
the pole contribution becomes dominant, and
the amplitude $t_{ij}$ obtained in the chiral unitary approach 
can be interpreted as the Breit-Wigner form,
\begin{equation}
    t_{ij}(\sqrt{s})
    \sim \frac{g_ig_j}{\sqrt{s}-M_{N^{*}}+i\Gamma_{N^{*}}/2}
    +t_{ij}^{BG} \ , \label{eq:BWresonance}
\end{equation}
where $M_{N^*}$ and $\Gamma_{N^*}$ are
the mass and width of the resonance, respectively.
Here the background term $t_{ij}^{BG}$ is 
assumed to be slowly varying function of $\sqrt{s}$,
and $g_i$ gives the coupling strength
of the resonance to the meson-baryon channel $i$.
A diagrammatic interpretation of Eq.~\eqref{eq:BWresonance}
is shown in Fig.~\ref{fig:MMamplitudes} (upper diagram).

In practical calculations,
we use two bases for the channels.
When we calculate the scattering amplitudes,
we adopt the physical basis such as $\pi^- p$ and $\eta n$,
because in the following subsections we introduce
the electromagnetic interactions,
which are not isospin symmetric.
While, when we calculate the resonance properties,
we change the basis to the isospin basis 
such as $\pi N(I=1/2)$ and $K\Sigma(I=3/2)$,
in order to specify the isospin of resonances.
The two bases are related each other through the Clebsh-Gordan
coefficients.

%%%%%%%%%%%%%%%%%%%%%%%%%%%%%%%%%%%%%%%%%%%%%%%%%%%%%%%%%%%%%%%%%%%%
\subsection{Electromagnetic interactions}\label{subsec:EMint}

By gauging the baryon kinetic term
in the lowest order $\mathcal{O}(p)$ Lagrangian,
we obtain the \( BB\gamma \) coupling,
which provides the normal magnetic moments of the 
ground state baryons.
In the effective chiral Lagrangian of order $\mathcal{O}(p^{2})$,
the photon coupling terms is given by~\cite{Meissner:1997hn}
\begin{equation}
    \begin{split} 
\mathcal{L}^{MB}_{(\gamma)}=&
-\frac{i}{4M_P}b_6^{F}
\Tr \left(
\bar{B}[S^{\mu},S^{\nu}]
[F^{+}_{\mu\nu},B]
\right) \\
&-\frac{i}{4M_P}b_6^{D}
\Tr \left(
\bar{B}[S^{\mu},S^{\nu}]
\{F^{+}_{\mu\nu},B\}
\right)
    \end{split}
    \label{eq:photonLag} \ ,
\end{equation}
where $M_P$ is the mass of proton, 
$b_6^F$ and $b_6^D$ are the low energy constants,
$F_{\mu\nu}^{+}=-e(\xi^{\dag}QF_{\mu\nu}\xi
+\xi QF_{\mu\nu}\xi^{\dag})$
with $\xi=\exp\{i\Phi/\sqrt{2}f\}$,
$Q=\text{diag}(2,-1,-1)/3$ and $S^{\mu}$ is a covariant
baryon spin operator.
$B$ and $\Phi$ are the octet baryon and meson fields, respectively.
This Lagrangian has contributions to anomalous magnetic moments.
Expanding the \( \xi \) fields,
we obtain the the magnetic moments of the ground state baryons
in terms of $b_6^D$ and $b_6^F$
in the chiral limit,
which satisfy the following Coleman-Glashow
relations~\cite{Coleman:1961jn};
\begin{equation}
    \begin{split}
\mu_{\Sigma^{+}}&=\mu_{p} \ ,\quad
2\mu_{\Lambda}=\mu_n \ ,\quad
\mu_{\Sigma^{-}}=\mu_{\Xi^{-}} \ , \\ 
\mu_{\Xi^{0}}&=\mu_{n} \ , \quad
\mu_{\Sigma^{-}}+\mu_n=-\mu_p \ , \\
2\mu_{\Sigma^{0}\Lambda}&=-\sqrt{3}\mu_n \ ,\quad
2\mu_{\Sigma^{0}}
=\mu_{\Sigma^{+}}+\mu_{\Sigma^{-}} \ .
    \end{split}
    \label{eq:CGrelations}
\end{equation}

Recall that the magnetic moments
derived from the Lagrangian \eqref{eq:photonLag}
are the anomalous magnetic moments,
while the normal magnetic moments come from
the covariant derivative.
However, the contributions from the normal magnetic moments
are exactly the same as the first term of Eq.~\eqref{eq:photonLag}.
Indeed the normal magnetic moments
just shift $b_6^{F}\to b_6^{F}+1$.
Therefore we will absorb the normal magnetic moments into
$b_6^{F}$ in the rest of this article,
in order to include the normal part into $b_6^F$ coefficients.
We need to be careful that
the values we show are different from 
the low energy constant $b_6^{F}$ which appears
in chiral perturbation theory.

Fitting the magnetic moments written in terms of
$b_6^{D}$ and $b_6^{F}$ to data,
we find the parameters~\cite{Meissner:1997hn}
\begin{equation}
    b_6^{D}=2.39,
    \quad b_6^{F}=1.77
    \label{eq:MMparameter}\ .
\end{equation}
In spite of the use of the only two parameters,
the tree level calculation
provides good results.

%%%%%%%%%%%%%%%%%%%%%%%%%%%%%%%%%%%%%%%%%%%%%%%%%%%%%%%%
\subsection{Photon coupling to resonances}\label{subsec:Pcouple}

\begin{figure}[tbp]
    \centering
    \includegraphics[width=8cm,clip]{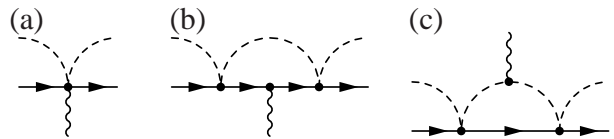}
    \caption{Photon coupling diagram in $-i\tilde{t}_{ij}(\sqrt{s})$.
    We consider that there are meson-baryon loops
    on the left and right sides of these vertices.}
    \label{fig:pcouple}
\end{figure}

Using the meson-baryon scattering amplitude $t_{ij}$ obtained 
in the chiral unitary approach and
electromagnetic couplings obtained in previous subsections,
here we calculate the photon coupling diagram
as shown in the bottom of Fig.~\ref{fig:MMamplitudes}.
In the chiral unitary model, where
the baryon resonances are expressed as
multiple scatterings of meson-baryon states,
we have three kinds of diagrams of the photon couplings
as shown in Fig.~\ref{fig:pcouple}~\cite{Jido:2002yz}.
Among them, the diagram (c) does not contribute to
the present calculation,
because we consider $s$ wave scattering.
Then, we write the amplitudes
$-i\tilde{t}_{ij}(\sqrt{s})$ as
\begin{equation}
    -i\tilde{t}_{ij}(\sqrt{s})
    =\left(-i\tilde{t}^{(a)}_{ij}(\sqrt{s})\right)+\left(
    -i\tilde{t}^{(b)}_{ij}(\sqrt{s})\right) \ ,
    \label{eq:Tdevide}
\end{equation}
where $-i\tilde{t}_{ij}^{(a)}(\sqrt{s})$ and
$-i\tilde{t}_{ij}^{(b)}(\sqrt{s})$
are the contributions
which includes the couplings 
of (a) and (b) in Fig.~\ref{fig:pcouple}.

The five point (two mesons, two baryons, and one photon)
vertex in Fig.~\ref{fig:pcouple} (a) is derived
from the Lagrangian~\eqref{eq:photonLag},
by taking the terms with two mesons.
We calculate the
amplitude of the tree vertex (a)
with these terms as
\begin{align}
    V^{BBMM\gamma}_{ij}
    &=ie\frac{\bm{\sigma}\times\mathbf{q}}
     {2M_p}\cdot\bm{\epsilon}A_{ij} \ , \nonumber
\end{align}
where
\begin{equation}
    A_{ij}=\frac{1}{2f^{2}}[X_{ij}b_6^{D}+Y_{ij}b_6^{F}] \ ,
    \label{eq:photoncouple}
\end{equation}
for convenience.
Coefficients $X_{ij}$ and $Y_{ij}$ are shown in
Tables~\ref{tbl:XYS0p0} and \ref{tbl:XYS0p1}.
\begin{table*}
    \caption{$X_{ij}$ and $Y_{ij}(S=0,Q=0)$}
    \label{tbl:XYS0p0}
\begin{ruledtabular}
    \begin{tabular}{ccccccc|cccccc}
& \multicolumn{6}{c|}{$X_{ij}$} 
& \multicolumn{6}{c}{$Y_{ij}$}  \\ 
& $\pi^{0}n$ & $\pi^{-}p$ & $\eta n$ & $K^{0}\Lambda$
& $K^{0}\Sigma^{0}$ & $K^{+}\Sigma^{-}$ 
& $\pi^{0}n$ & $\pi^{-}p$ & $\eta n$ & $K^{0}\Lambda$
& $K^{0}\Sigma^{0}$ & $K^{+}\Sigma^{-}$ \\ \hline
    $\pi^{0}n$ & 0 & $\frac{1}{\sqrt{2}}$ & 0 & 0
& 0 & $\frac{1}{2\sqrt{2}}$
& 0 & $\frac{1}{\sqrt{2}}$ & 0 & 0
& 0 & $-\frac{1}{2\sqrt{2}}$ \\
    $\pi^{-}p$ & & $-1$ & 0 & $\frac{1}{2\sqrt{6}}$
& $-\frac{1}{2\sqrt{2}}$ & 0
& & $-1$ & 0 & $\frac{\sqrt{6}}{4}$
& $\frac{1}{2\sqrt{2}}$ & 0 \\ 
    $\eta n$ & & & 0 & 0
& 0 & $\frac{\sqrt{6}}{4}$
& & & 0 & 0
& 0 & $-\frac{\sqrt{6}}{4}$ \\ 
    $K^{0}\Lambda$ & & & & 0 & 0 & $-\frac{1}{\sqrt{6}}$ 
& & & & 0 & 0 & 0 \\
    $K^{0}\Sigma^{0}$ & & & & & 0 & 0 
& & & & & 0 & $-\frac{1}{\sqrt{2}}$ \\
    $K^{+}\Sigma^{-}$ & & & & & & $-1$ 
& & & & & & 1 \\ 
    \end{tabular}
\end{ruledtabular}
\vspace*{0.3cm}
\centering
\caption{$X_{ij}$ and $Y_{ij}(S=0,Q=1)$}
\label{tbl:XYS0p1}
\begin{ruledtabular}
    \begin{tabular}{ccccccc|cccccc}
& \multicolumn{6}{c|}{$X_{ij}$} 
& \multicolumn{6}{c}{$Y_{ij}$}  \\ 
& $\pi^{0}p$ & $\pi^{+}n$ & $\eta p$ & $K^{+}\Lambda$
& $K^{+}\Sigma^{0}$ & $K^{0}\Sigma^{+}$ 
& $\pi^{0}p$ & $\pi^{+}n$ & $\eta p$ & $K^{+}\Lambda$
& $K^{+}\Sigma^{0}$ & $K^{0}\Sigma^{+}$ \\ \hline
$\pi^{0}p$ & $0$ & $\frac{1}{\sqrt{2}}$ & 0 
& $-\frac{1}{4\sqrt{3}}$ & $\frac{1}{4}$ & 0 
& $0$ & $\frac{1}{\sqrt{2}}$ & 0 
& $-\frac{\sqrt{3}}{4}$ & $-\frac{1}{4}$ & 0  \\ 
$\pi^{+}n$ & & $1$ & 0 & $-\frac{1}{\sqrt{6}}$
& $-\frac{1}{\sqrt{2}}$  & 0
& & $1$ & 0 & $-\sqrt{\frac{3}{2}}$
& $\frac{1}{\sqrt{2}}$ & 0 \\ 
$\eta p$ & & & 0 & $-\frac{1}{4}$
& $\frac{\sqrt{3}}{4}$ & 0 
& & & 0 & $-\frac{3}{4}$
& $-\frac{\sqrt{3}}{4}$ & 0 \\ 
$K^{+}\Lambda$ & & & & 1 & $-\frac{1}{\sqrt{3}}$
& $-\frac{1}{\sqrt{6}}$ 
& & & & 0 & $0$ & 0 \\
$K^{+}\Sigma^{0}$ & & & & & $-1$ & 0 
& & & & & $0$ & $\frac{1}{\sqrt{2}}$ \\
$K^{0}\Sigma^{+}$ & & & & & & 0 & & & & & & 0\\
    \end{tabular}
\end{ruledtabular}
\end{table*}
In order to calculate the resonance magnetic moments,
we attach %(*** or other word ***) 
the meson-baryon scattering amplitude
obtained in the chiral unitary model
to the both side of the diagram (a) in Fig.~\ref{fig:pcouple}.
We then obtain
\begin{equation}
    -i\tilde{t}^{(a)}_{ij}
    =t_{il}G_{l}A_{lm}G_mt_{mj} \ .
    \label{eq:Ta}
\end{equation}

The diagram (b) in Fig.~\ref{fig:pcouple} is
calculated by magnetic moments of the ground state baryons $\mu_i$
multiplying by a loop function with the two baryon propagators.
In the limit that the photon momentum goes to zero,
this loop function reduces to a simple form as
\begin{align}
    &\tilde{G}_{i}(\sqrt{s}) \nonumber \\
    = & i\int\frac{d^{4}q}{(2\pi)^{4}}
    \frac{2M_i}{(P-q)^{2}-M_i^{2}}
    \frac{2M_i}{(P-q)^{2}-M_i^{2}}
    \frac{1}{q^{2}-m_i^{2}} 
    \nonumber \\
    =&-\frac{\partial}{\partial \sqrt{s}}G_i(\sqrt{s})
    \label{eq:Gtilde}  \ .
\end{align}
This analytic form of $\tilde{G}_{i}(\sqrt{s})$
is convenient when we search the poles of the amplitude
in the complex plane,
which correspond to resonances.
Finally we obtain
\begin{equation}
    -i\tilde{t}^{(b)}_{ij}
    =t_{il}\tilde{G}_{l}\mu_{l}t_{lj}
    \ , \label{eq:Tb}
\end{equation}
where $\mu_{l}$ are the magnetic moments of the ground state baryons.

\begin{figure}[tbp]
    \centering
    \includegraphics[width=6cm,clip]{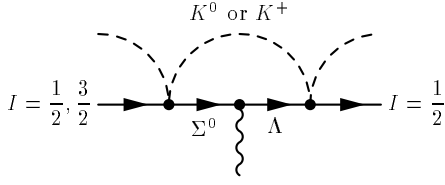} \\
    \caption{Diagrams of off-diagonal components in 
    $\tilde{G}$ including $\Sigma^{0}\Lambda$ transition
    in the $S=0$ channel.}
    \label{fig:SLtrans}
\end{figure}

Since there is the $\Sigma^{0}\Lambda$ transition
moment in the ground state,
off-diagonal components exist in $\tilde{G}$
as shown in Fig.~\ref{fig:SLtrans}.
In order to take it into account,
we need further approximation,
because the masses in the first and second propagators
in Eq.~\eqref{eq:Gtilde} are different.
We calculate these off-diagonal components by taking
average of the $\Sigma^{0}$ and $\Lambda$ propagators,
namely,
\begin{equation}
    \tilde{G}_{M\Lambda,M\Sigma^0}(\sqrt{s})
    =\frac{1}{2}\left(
    \tilde{G}_{M\Sigma^{0}}(\sqrt{s})
    +\tilde{G}_{M\Lambda}(\sqrt{s})\right) \ ,
    \label{eq:offdiag}
\end{equation}
where $M$ denotes $K^{0}$ or $K^{+}$.
Then Eq.~\eqref{eq:Tb} is modified as
\begin{equation}
    -i\tilde{t}^{(b)}_{ij}
    =t_{il}
    \bigl[\delta_{lm}\tilde{G}_{l}\mu_{l}
    +\tilde{G}_{l,m}
    \, \mu_{(\Sigma^0\Lambda)}
    \bigr]t_{mj}
    \ , \label{eq:Tb2}
\end{equation}
where $\tilde{G}_{l,m}\, \mu_{(\Sigma^0\Lambda)}$
represents the corresponding 
transition channels as shown in Fig.~\ref{fig:SLtrans}.

As pointed out in Ref.~\cite{Jido:2002yz},
the effects of $\Sigma^0\Lambda$ transition is 
almost negligible in the $S=-1$ channel,
where the corresponding processes are
$\pi^0\Sigma^0\to \pi^0\Lambda$ and $\eta\Sigma^0\to \eta\Lambda$.
These transition terms do not contribute to the magnetic moments
of $\Lambda$ resonances ($I=0$) in the isospin limit.
However, for the $N$ resonance they do.
The $\Sigma^0\Lambda$ transition
occurs among the $K\Sigma_0$ and $K\Lambda$ channels of $I=1/2$.
After projecting them to $I=1/2$ state,
we have checked numerically that
the inclusion of the $\Sigma^0\Lambda$ transition
has a moderate effect to the amplitude $-i\tilde{t}^{(b)}_{ij}$.

In this way, we obtain the amplitude $-i\tilde{t}$
through Eqs.~\eqref{eq:Tdevide}, \eqref{eq:Ta} and \eqref{eq:Tb2}
in the chiral unitary model.
Around the resonance energy region, 
where the pole contribution becomes dominant,
$-i\tilde{t}$ obtained in the chiral unitary model
can be interpreted using
the Breit-Wigner form of resonance:
\begin{equation}
    \begin{split}
-i\tilde{t}_{ij}(\sqrt{s})
\sim&
\left(\frac{g_i}{\sqrt{s}-M_{N^{*}}+i\Gamma_{N^{*}}/2}
+t^{BG}\right) \\
&\cdot
\mu_{N^{*}} \cdot 
\left(\frac{g_j}{\sqrt{s}-M_{N^{*}}+i\Gamma_{N^{*}}/2}
+t^{BG}\right)\ ,
    \end{split}
    \label{eq:ttilde}
\end{equation}
where $\mu_{N^{*}}$ is the magnetic moment
of the $N^*$ resonance.

%%%%%%%%%%%%%%%%%%%%%%%%%%%%%%%%%%%%%%%%%%%%%%%%%%%%%%%%

\subsection{Evaluation of the magnetic
moments}\label{subsec:evaluation}

Here we evaluate the magnetic moments of the
resonances from the amplitudes $t_{ij}$ and $-i\tilde{t}_{ij}$
obtained by the chiral unitary model.
Around the resonance energy region $\sqrt{s}\sim M_{N^*}$, 
we can regard these amplitudes as
Eqs.~\eqref{eq:BWresonance} and \eqref{eq:ttilde}, respectively.
Using $t_{ij}$, we eliminate the resonance propagators 
and couplings $g_i$ in $-i\tilde{t}_{ij}$.
There are two methods,
one is to extract them on the real axis 
and the other is to evaluate them in the complex plane \cite{Jido:2002yz}.

On the real axis, in the resonance dominance, where background terms are neglected,
the magnetic moment can be evaluated by
\begin{equation}
    \mu_{N^*}(\sqrt{s}) \sim
    \frac{-i\tilde{t}_{ij}(\sqrt{s})}
    {-\frac{\partial}{\partial \sqrt{s}}t_{ij}(\sqrt{s})} \ ,
    \label{eq:mureal1}
\end{equation}
where the denominator cancels the two resonance propagators
and couplings, leaving the factor of magnetic moment.
In actual cases, however, there are background contributions,
which we have to deal with carefully.
To show explicitly, the amplitude $-i\tilde{t}_{ij}$
in Eq.~\ref{eq:mureal1} with the background term included,
should be divided by the factor
\begin{equation}
    \frac{\partial}{\partial \sqrt{s}}t_{ij}(\sqrt{s})
    =-\frac{g_ig_j}{(\sqrt{s}-M_{N^{*}}+i\Gamma_{N^{*}}/2)^{2}}
    +\frac{\partial}{\partial \sqrt{s}}t_{ij}^{BG}
    \label{eq:difT2} \ .
\end{equation}
However, since the background term $t_{ij}^{BG}$
is assumed to be a slowly varying function of $\sqrt{s}$,
its derivative must be small,
therefore we can neglect it.
The result is
\begin{align}
    \frac{-i\tilde{t}_{ij}(\sqrt{s})}
    {-\frac{\partial}{\partial \sqrt{s}}t_{ij}(\sqrt{s})}
    = &\mu_{N^{*}}(\sqrt{s})
    \nonumber \\
    &+t^{BG}
    \frac{\sqrt{s}-z_{N^*}}{g_i}
    +t^{BG}
    \frac{\sqrt{s}-z_{N^*}}{g_j}\nonumber \\
    &+\left(t^{BG}\right)^{2}
    \frac{(\sqrt{s}-z_{N^*})^{2}}{g_ig_j}
    \label{eq:BGeffects} \ ,
\end{align}
where $z_{N^*}\equiv M_{N^{*}}-i\Gamma_{N^{*}}/2$.
Note that the second and third lines in Eq.~\eqref{eq:BGeffects}
are not always regarded as small,
because $z_{N^*}$ has an imaginary part, so that
$(\sqrt{s}-z_{N^*})$ cannot be zero.
In order to make these background terms small,
we evaluate $\mu_{N^{*}}$ 
at $\sqrt{s}\sim M_{N^*}$ and choose the suitable
channel which strongly couples to the resonance.
%in the channel whose coupling $|g_i|^{2}$ is large.

The analytic form of the amplitude
enable us to calculate the magnetic moment in the complex plane.
It is a big advantage of the chiral unitary model, 
since we can calculate the magnetic moment exactly on the pole
and, therefore, the background terms do not give any contributions.
At the resonance pole $z\to z_{N^*}$,
we calculate 
\begin{align}
    &\lim_{z\to z_{N^*}}(z-z_R)\frac{-i\tilde{t}_{ij}(z)}
    {t_{ij}(z)} 
    \nonumber \\
    =&\lim_{z\to z_{N^*}}
    \left[
    \frac{\mu_{N^{*}}(z)}{1+(z-z_{N^*})t^{BG}/(g_ig_j)}
    +\mathcal{O}(z-z_{N^*})
    \right] \nonumber \\
    =&\mu_{N^{*}}(z_{N^*}) \label{eq:comMM} \ .
\end{align}
%where the background terms vanish,
%because we can calculate it exactly on the pole.
Since the position of the pole generated in the unitarization
does not depend on the channel,
the result \eqref{eq:comMM} is independent of the channel
chosen to calculate the magnetic moments.

In the second method,
we can compute $\mu_{N^*}(z_{N^*})$ without ambiguity,
but with a complex phase.
Because of this,
we discuss only the absolute value $|\mu_{N^*}(z_{N^*})|$.
On the other hand, in the first method,
we can determine the sign,
although the background terms
make the absolute values ambiguous.
We can minimize such ambiguities by
choosing the most relevant channel,
where the coupling strength of the pole $g_i$ 
is the largest.

Hence, our strategy here is to
calculate the absolute value in the complex plane
and determine the sign on the real axis.
It is not trivial that
$|\mu_{N^*}(z_R)|$ and $|\mu_{N^*}(\sqrt{s})|$ take a same value,
but we expect that
they should be close each other.

%%%%%%%%%%%%%%%%%%%%%%%%%%%%%%%%%%%%%%%%%%%%%%%%%%%%%%%%%%%%%%%%%%%%

\section{Numerical results}\label{sec:Nresults}

In this section, we show the results
of numerical calculations.
First, we present the input parameters and
calculate $S=0$ meson-baryon scatterings
in the chiral unitary model.
Next, using the same parameters,
we calculate the magnetic moments
of the $N(1535)$ resonance.
In the following,
we denote the two charge states of $N(1535)$
as $n^*(Q=0)$ and $p^*(Q=1)$.

%%%%%%%%%%%%%%%%%%%%%%%%%%%%%%%%%%%%%%%%%%%%%%%%%%%%%%%%%%%%%%%%%%%%
\subsection{The $N(1535)$ resonance 
in the chiral unitary model}\label{subsec:Nres}

The resonance states are obtained by
solving the scattering equation~\eqref{eq:Tmat},
whose coefficients
$C_{ij}$ are shown in Table~\ref{tbl:Cij}.
The coefficients $X_{ij}$ and $Y_{ij}$ 
for the magnetic moments are given in Tables~\ref{tbl:XYS0p0} and
\ref{tbl:XYS0p1}.
For the mass of the particles $m_i$ and $M_i$,
and the magnetic moments of the ground state baryons $\mu_i$,
we use the physical values
taken from the Particle Data Group (PDG)~\cite{Hagiwara:2002fs}.
The low energy constants $b_6^F$ and $b_6^D$
are given by Eq.~\eqref{eq:MMparameter}.
In order to calculate the loop function~\eqref{eq:loop},
we use the regularization scale $\mu=630$ MeV 
and the following channel dependent subtraction constants
taken from Ref.~\cite{Inoue:2001ip};
\begin{equation}
    \begin{split}
a_{\pi N}&=0.711 \ , \quad 
a_{\eta N}=-1.09 \ , \\
a_{K\Lambda}&=0.311 \ , \quad 
a_{K\Sigma}=-4.09 \ .
    \end{split}
    \label{eq:subtractions}
\end{equation}
Here we show the shifted values of $a_i$ corresponding 
to $\mu=630$ MeV  by using the relation 
$a(\mu^{\prime})=a(\mu)+2\ln(\mu^{\prime}/\mu)$.
We use the common value for each isospin multiplet.
These constants are essential to generate 
the $N(1535)$ resonance~\cite{Hyodo:2002pk,Hyodo:2003}.
We adopt the physical meson decay constants,
\begin{equation}
    f_{\pi}=93 \ \ \text{[MeV]}\ ,
    \quad f_{K}=1.22\times f_{\pi} \ ,
    \quad f_{\eta}=1.3\times f_{\pi} \ ,
    \label{eq:fmeson}
\end{equation}
following Ref.~\cite{Inoue:2001ip}.

Using these inputs,
we calculate the scattering amplitudes~\eqref{eq:Tmat},
which well describes the $S_{11}$ phase shifts,
the scattering amplitudes and the total cross section
of $\pi^-p\to \eta n$.
In the complex energy plane, we find poles at
\begin{equation}
    \begin{split}
z_{n^{*}}&= 1536.01
-37.06i \ \ \text{[MeV]}
\quad (Q=0) \ , \\
z_{p^{*}}&= 1531.01
-36.38i \ \ \text{[MeV]}
\quad (Q=1) \ ,
    \end{split}
    \label{eq:Npoles}
\end{equation}
whose real and imaginary parts correspond
to the mass $M_{N^*}$ and width $\Gamma_{N^*}/2$, respectively,
for the Breit-Wigner parametrization~\eqref{eq:BWresonance}.
Note that before including the electromagnetic interactions,
the difference between $n^*$ and $p^*$
comes from the tiny isospin violation due to the particle mass differences.
Their coupling strengths $|g_i|^2$
to the various meson-baryon channels are shown
in Table~\ref{tbl:Ncouplings}.
\begin{table}[tbp]
    \caption{Coupling strengths of the $N(1535)$ resonance
    to meson-baryon channels.
    All the channels have isospin $I=1/2$.}
    \label{tbl:Ncouplings}
    \centering
    \begin{ruledtabular}
    \begin{tabular}{ccccc}
     & $|g_{\pi N}|^{2}$ & $|g_{\eta N}|^{2}$
     & $|g_{K\Lambda}|^{2}$ & $|g_{K\Sigma}|^{2}$  \\
    \hline
    $n^{*}$ & 0.623 & 2.30 & 1.93 & 7.29  \\
    $p^{*}$ & 0.619 & 2.35 & 1.88 & 7.37 
    \end{tabular}
\end{ruledtabular}
\end{table}
>From this table, we see that the $K\Sigma$ channel
has the largest coupling strength to the $N(1535)$ resonance,
which indicates that the resonance is a
quasi-bound state of $K\Sigma$,
as pointed out in Ref.~\cite{Kaiser:1995cy}.
At the energy of the threshold of $N(1535)$,
the $\pi N$ and $\eta N$ channels open.
Therefore, the decay of the resonance is dominated by
$\eta N$ channel, which is a
characteristic properties of $N(1535)$.

%%%%%%%%%%%%%%%%%%%%%%%%%%%%%%%%%%%%%%%%%%%%%%%%%%%%%%%%%%%%%%%%%%%%

\subsection{Magnetic moments of
the $N(1535)$ resonance}\label{subsec:Mmoment}

We first show the results in the complex plane,
calculated from Eq.~\eqref{eq:comMM}.
The results for the absolute values are
\begin{equation}
    |\mu_{n^{*}}|
    =0.248\mu_N \ , \quad
    |\mu_{p^{*}}|
    =1.13\mu_N \ .
    \label{eq:Cmag}
\end{equation}
As expected, we have checked 
that the results are channel independent,
since we can eliminate the background contributions.

\begin{figure*}[tbp]
    \includegraphics[width=16cm,clip]{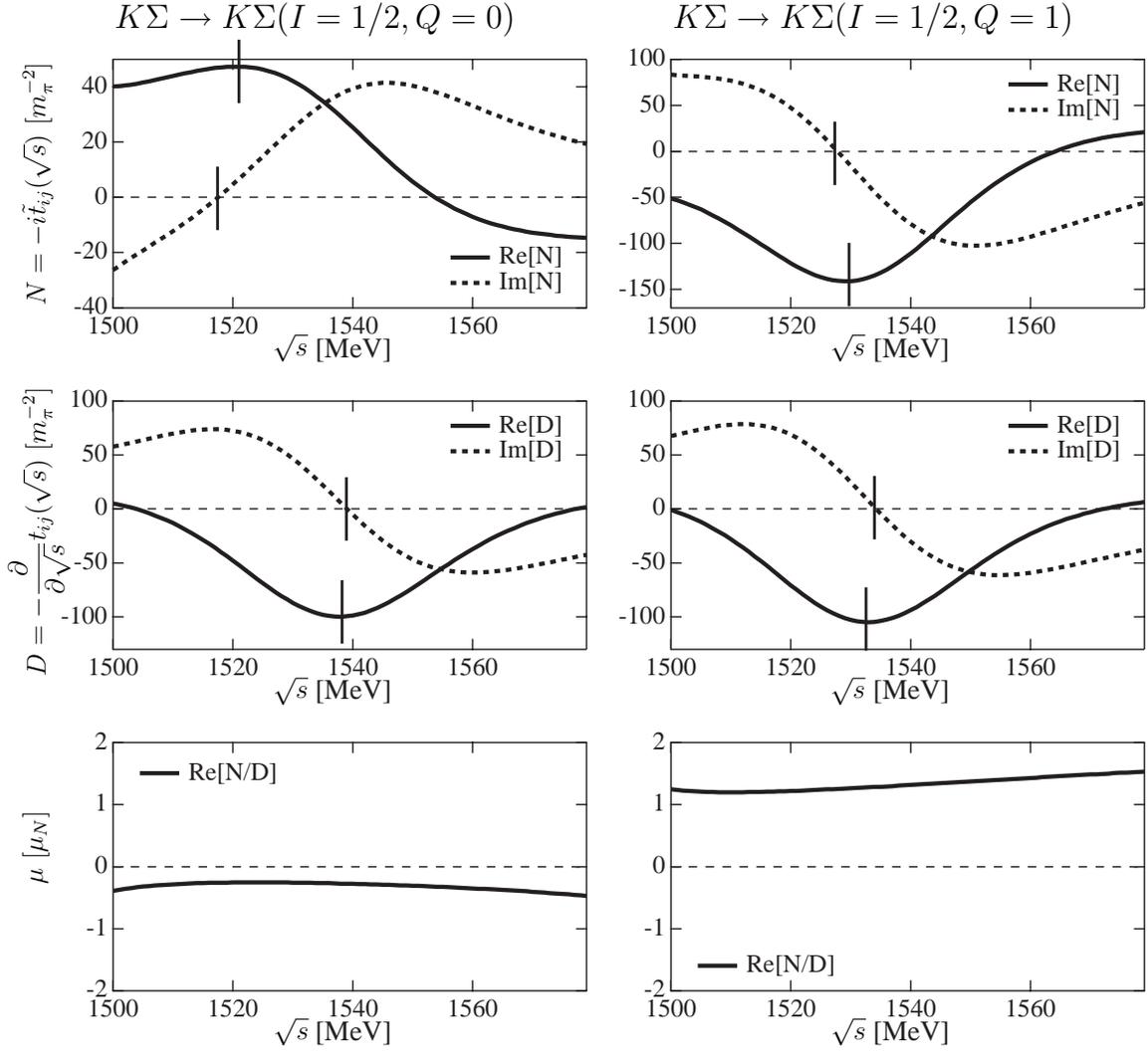}
    \caption{
    Scattering amplitudes and
    the magnetic moments on the real axis.
    We plot the real and imaginary parts
    of the $K\Sigma\to K\Sigma$
    amplitudes $N=-i\tilde{t}_{ij}(\sqrt{s})$
    and $D=-\frac{\partial}{\partial\sqrt{s}}t_{ij}(\sqrt{s})$,
    in \( Q=0 \) and $Q=1$.
    Solid bars represent the position $\sqrt{s}=M_R$,
    expected by the Breit-Wigner form.
    The magnetic moments $\mu=\re[N/D]$
    are calculated in units of the nuclear magneton.}
    \label{fig:Rmag}
\end{figure*}

Next we calculate $\mu(\sqrt{s})$ on the real axis.
In Fig.~\ref{fig:Rmag},
we plot the scattering amplitudes
$N=-i\tilde{t}_{ij}(\sqrt{s})$
and $D=-\frac{\partial}{\partial\sqrt{s}}t_{ij}(\sqrt{s})$,
and the magnetic moments $\mu\sim N/D$.
Here we plot the amplitudes
$K\Sigma\to K\Sigma(I=1/2)$,
since it has the largest coupling strength to the resonance
and background contributions are expected to be small.
Notice that 
the amplitudes $D$ do not
contain the electromagnetic interaction,
and therefore,
$D$ for $Q=0$ and $Q=1$
are the same when isospin
violation is neglected.
On the other hand, the amplitudes $N$
are different from each other due to the photon couplings.
It is important that
$N(Q=0)$ and $N(Q=1)$
have opposite signs.
These signs together with $D$
determines the sign of the magnetic moments.

If there were no background,
for both $N$ and $D$,
extreme values of the real parts
and zeros of the imaginary parts
would have taken place at the same value $\sqrt{s}=M_{N^*}$.
In this case, $\mu=N/D$ becomes pure real.
However, in actual calculations
these points deviate slightly,
especially in $Q=0$,
due to background contributions.
Therefore, we evaluate the magnetic moments at all these points.
Since $\mu=N/D$ has a small imaginary part
due to the backgrounds,
we calculate $\mu$ as,
\begin{equation}
    \mu=\re\left[\frac{-i\tilde{t}_{ij}(\sqrt{s})}
    {-\frac{\partial}{\partial\sqrt{s}}
    t_{ij}(\sqrt{s})}\right] \ .
    \label{eq:remucal} 
\end{equation}
The results are shown in the bottom figures of Fig.~\ref{fig:Rmag}
Then we determine the value of the magnetic moments 
of the resonance
\begin{equation}
    \begin{split}
\mu_{n^{*}}
&=(-0.266\pm 0.01)\mu_N \ ,  \\
\mu_{p^{*}}
&=(1.26\pm 0.02)\mu_N \ ,
    \end{split}
    \label{eq:Rmag}
\end{equation}
with small uncertainties.
The absolute values of Eq.~\eqref{eq:Rmag}
do not differ very much
from the results~\eqref{eq:Cmag}.
This is because we adopt the $K\Sigma$ channel,
where $|g_{i}|^2$ is the largest.
When we choose the other channels
to evaluate the magnetic moments,
the difference from the results~\eqref{eq:Cmag} becomes larger,
due to the background effects
(second and third lines of Eq.~\eqref{eq:BGeffects}).

Finally we summarize the results in Table~\ref{tbl:sumtable}.
\begin{table}[bp]
    \caption{The magnetic moments of the $N(1535)$ resonance
in units of the nuclear magneton.}
\label{tbl:sumtable}
    \centering
\begin{ruledtabular}    
    \begin{tabular}{lll}
     & $\phantom{-}n^{*}$ & $p^{*}$  \\
     \hline
     $|\mu|$ (complex plane) & $\phantom{-}0.248$ & 1.13  \\
    $\phantom{|}\mu\phantom{|}$  (real axis)
    & $-0.266\pm 0.01$ & $1.26\pm 0.02$   \\
    \end{tabular}
\end{ruledtabular}
\end{table}
Combining the results in the complex plane and on the real axis,
we determine the magnetic moments as
\begin{equation}
    \mu_{n^{*}(1535)}= -0.25\mu_N \ ,\quad
     \mu_{p^{*}(1535)}= 1.1\mu_N \ .
    \label{eq:results}
\end{equation}
In next section,
we discuss these results in detail.

%%%%%%%%%%%%%%%%%%%%%%%%%%%%%%%%%%%%%%%%%%%%%%%%%%%%%%%%%%%%%%%%%%%%
\section{Discussions}\label{sec:discussion}

First we discuss the SU(3) relation
by comparing the present results with
the magnetic moment of $\Lambda(1670)$
obtained in the same framework~\cite{Jido:2002yz}.
Then we decompose the magnetic moment
into the various components in order to understand the origin of
the obtained values.
Then we discuss the magnetic moments in the quark model
and chiral doublet model, in comparison with the present results.

\subsection{The SU(3) relation}

In Ref.~\cite{Jido:2002yz},
the magnetic moments of $\Lambda(1670)$ are
calculated in the chiral unitary model;
\begin{equation}
    \mu_{\Lambda^*(1670)}= -0.29\mu_N \ .
    \label{eq:ResMoments}
\end{equation}
The $\Lambda(1670)$ and $N(1535)$
have $J^P=1/2^{-}$, so that they have been 
considered to be members of the SU(3) octet.
If the SU(3) symmetry is exact,
the magnetic moments of the octet should satisfy
the Coleman-Glashow relations
in Eq.~\eqref{eq:CGrelations},
which tell us that
\begin{equation}
    \mu_{n^{*}}=2\mu_{\Lambda^{*}} \ .
    \label{eq:CGrelation2}
\end{equation}
In the present calculation,
the signs of the magnetic moments in
Eqs.~\eqref{eq:Rmag}~and~\eqref{eq:ResMoments}
are consistent with this relation,
although the absolute values do not agree with
Eq.~\eqref{eq:CGrelation2}.

The SU(3) relation is discussed more clearly by
looking at the SU(3) decomposition of the resonance states
in terms of  the coupling strengths $g_i$.
The coupling strengths in the SU(3) basis
are obtained by a unitary transformation
using SU(3) Clebsh-Gordan coefficients~\cite{Jido:2003cb}.
In Table~\ref{tbl:SU3},
\begin{table}[tbp]
    \caption{Coupling strengths 
    $|g_{i}|^{2}$ of $N(1535)$ and $\Lambda(1670)$
    in SU(3) basis.
    Values for $\Lambda^*(1670)$ are taken from
    Ref.~\cite{Jido:2003cb}}
    \label{tbl:SU3}
    \centering
    \begin{ruledtabular}
    \begin{tabular}{ccccccc}
     representation & $1$ & $8$ & $8$ & $10$ 
     & $\bar{10}$ & $27$  \\
    \hline
    $n^{*}(1535)$ & $-$ &  5.2 & 6.2 & 0.17 & $-$ & 0.58  \\
    $\Lambda^{*}(1670)$ & 4.0 & 2.3 & 7.3 & $-$ & $-$ &  0.16 
    \end{tabular}
\end{ruledtabular}
\end{table}
$|g_{i}|^{2}$ in SU(3) basis are shown,
where we observe that for both $N(1535)$ and $\Lambda(1670)$,
octet component are dominant.
This fact explains
qualitative agreement
of the relation between $\mu_{n^*}$ and $\mu_{\Lambda^*}$
in the chiral unitary model.
Deviation from the relation
comes from the 
large mixture of the singlet component in $\Lambda(1670)$
and SU(3) breaking effects.

\subsection{Isospin decomposition}

For later discussions, we decompose the magnetic moments
in Eq.~\eqref{eq:results}
into isoscalar ($\mu_S$) and isovector ($\mu_V$) components.
These moments are defined by
\begin{equation}
     \mu_{
     \genfrac{}{}{0pt}{}{S}{V}}
     =\frac{1}{2}(\mu_p\pm \mu_n) \ .
     \label{eq:isodef}
\end{equation}
In units of nuclear magneton $\mu_N=e/2M_N$,
these values are
\begin{equation}
    \mu_{S}= 0.44\mu_{N} \ ,\quad
    \mu_{V}= 0.69\mu_{N} \ ,
    \label{eq:isobasis1}
\end{equation}
The isoscalar magnetic moment of $N(1535)$
is similar to that of the ground state nucleon $N(939)$,
but the isovector one is much smaller than
that of the nucleon $\mu_{V}(939)=2.35\mu_N$.

More quantitatively, it is considered to express
these values in units of resonance magneton
$\mu_{N^*}\equiv e/2M_{N^*}$
and extract the anomalous magnetic moments $\kappa$
in units of $\mu_{N^*}$.
The results are
\begin{equation}
    \mu_{S}(1535)= 0.72\mu_{N^*} \ ,\quad
    \mu_{V}(1535)= 1.13\mu_{N^*} \ ,
    \label{eq:isobasis}
\end{equation}
and 
\begin{equation}
    \kappa_{S}(1535)= 0.22\mu_{N^*} \ ,\quad
    \kappa_{V}(1535)= 0.63\mu_{N^*} \ .
    \label{eq:isobasis2}
\end{equation}
These numbers may be compared with those of the nucleon
(in units of nuclear magneton):
\begin{equation}
    \kappa_{S}(939)= -0.06\mu_{N} \ ,\quad
    \kappa_{V}(939)= 1.85\mu_{N} \ .
    \label{eq:isobasis3}
\end{equation}
Hence the strong isovector dominance
as in the $N(939)$ magnetic moments
is not realized in $N(1535)$.

\subsection{Contributions from various terms}

\begin{figure*}[tbp]
    \includegraphics[width=14cm,clip]{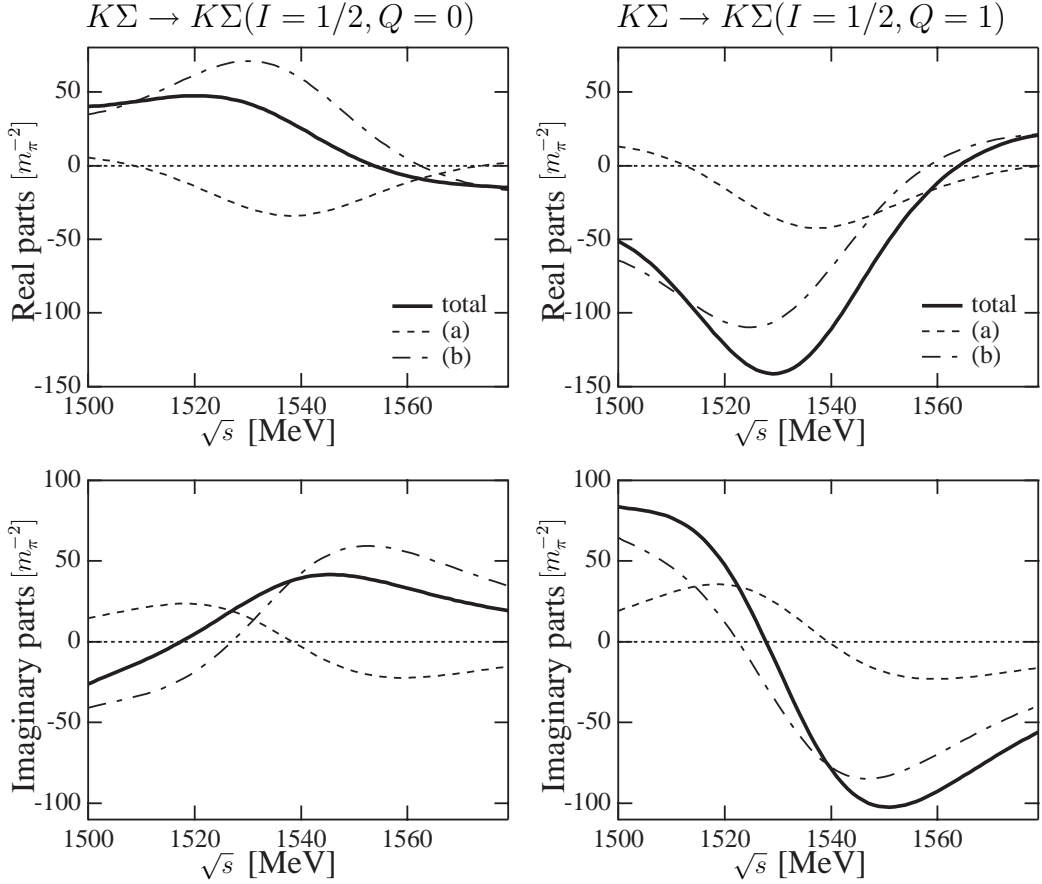}
    \caption{Real and imaginary parts
    of the $K\Sigma\to K\Sigma$
    amplitudes $-i\tilde{t}^{(a)}_{ij}$,
    $-i\tilde{t}^{(b)}_{ij}$ and $-i\tilde{t}_{ij}
    =-i\tilde{t}^{(a)}_{ij}-i\tilde{t}^{(b)}_{ij}$
    (dashed, dash-dotted and solid lines)
    in $Q=0$ and $Q=1$.}
    \label{fig:cont}
\end{figure*}

In this subsection,
we decompose the magnetic moments into various terms
in order to understand qualitatively their origins.
First we decompose the amplitude 
into the contributions from
the term in Fig.~\ref{fig:pcouple} (a) (contact vertex)
and those from in Fig.~\ref{fig:pcouple} (b)
(photon attached to baryon propagator).
We perform this decomposition for the
results obtained on the real axis.
In Fig.~\ref{fig:cont},
we show the corresponding amplitudes 
for $Q=0$ and $Q=1$.
We see that the contribution from (a)
is smaller in magnitude than (b),
and that the contributions of (a) and (b)
have opposite (same) signs for $Q=0$ ($Q=1$).
Therefore, there is a cancellation between them
for $Q=0$ in the total value, 
while for $Q=1$ two terms are added
with the same sign.
This explains partly smaller magnetic moments of $Q=0$
than that of $Q=1$.
The actual numbers around resonance region are given as
\begin{equation}
\begin{split}
    \mu^{(a)}_{n^*}&\sim 0.34\mu_N \ , \quad
    \mu^{(b)}_{n^*}\sim -0.60\mu_N \ , \\
    \mu^{(a)}_{p^*}&\sim 0.40\mu_N \ , \quad
    \mu^{(b)}_{p^*}\sim 0.87\mu_N \ ,
\end{split}
\end{equation}
where $\mu_{n^*,p^*}=\mu^{(a)}_{n^*,p^*}+\mu^{(b)}_{n^*,p^*}$.
As compared with Eq.~\eqref{eq:isobasis1},
these numbers imply that the term $\mu^{(a)}$ 
is dominated by the isoscalar piece,
while the term $\mu^{(b)}$ 
by the isovector piece.
The nonegligible values of $\mu^{(a)}$ 
is the origin to weaken isovector dominance of the
$N(1535)$ magnetic moments.

Let us now consider the the piece $\mu^{(b)}$
and its isovector dominance.
As shown in Fig.~\ref{fig:pcouple} (b), 
$\mu^{(b)}$ is given by a sum of the diagrams 
where the photon couples to a ground state baryon.
In this case, we can draw a naive picture
where the resonance magnetic moment can be written as
a sum of the magnetic moments of the ground state baryons
weighted by their probabilities in the resonance wave function.
In Ref~\cite{Chiang:2002ah},
considering $N(1535)$ as a quasi-bound state of $K\Sigma$,
they decomposed the $K\Sigma$ isospin state into physical states
by the Clebsh-Gordan coefficients,
and evaluate the magnetic moments of the ground state baryon.
Here we extend this estimation to
sum up all the channels,
multiplying $|g_i|^{2}$ as weight.
First, we define the magnetic moments
of the isospin states,
using the Clebsh-Gordan coefficients
and the magnetic moments of the ground states;
\begin{equation}
    \begin{split}
\mu_{\pi N}(Q=0)&=\frac{1}{3}\mu_{n}
+\frac{2}{3}\mu_{p}\sim 1.22\mu_N \ ,\\
\mu_{\eta N}(Q=0)&=\mu_{n}\sim -1.91\mu_N \ ,\\
\mu_{K\Lambda}(Q=0)&=\mu_{\Lambda}\sim -0.613\mu_N \ ,\\
\mu_{K\Sigma}(Q=0)&=\frac{1}{3}\mu_{\Sigma^{0}}
+\frac{2}{3}\mu_{\Sigma^{-}}\sim -0.557\mu_N \ ,\\
\mu_{\pi N}(Q=1)&=\frac{2}{3}\mu_{n}
+\frac{1}{3}\mu_{p}\sim -0.343\mu_N \ ,\\
\mu_{\eta N}(Q=1)&=\mu_{p}\sim 2.79\mu_N\ ,\\
\mu_{K\Lambda}(Q=1)&=\mu_{\Lambda}\sim -0.613\mu_N \ ,\\
\mu_{K\Sigma}(Q=1)&=\frac{1}{3}\mu_{\Sigma^{0}}
+\frac{2}{3}\mu_{\Sigma^{+}}\sim 1.86\mu_N \ .
    \end{split}
\end{equation}
Multiplying the weight $|g_i|^{2}$
in Table~\ref{tbl:Ncouplings},
we calculate
%%%%%%%%%%%%%%%%%%%% Eq. (37) %%%%%%%%%%%%%%%%%%%%%%%%%%%%%%%%%%%
% 
% Is it ok ?
\begin{equation}
    \begin{split}
\mu_{N^{*}}
=&\frac{|g_{\pi N}|^{2}}{\sum_{j}|g_j|^{2}}
\mu_{\pi N}
+\frac{|g_{\eta N}|^{2}}{\sum_{j}|g_j|^{2}}
\mu_{\eta N} \\
&+\frac{|g_{K\Lambda}|^{2}}{\sum_{j}|g_j|^{2}}
\mu_{K\Lambda}
+\frac{|g_{K\Sigma}|^{2}}{\sum_{j}|g_j|^{2}}
\mu_{K\Sigma}
    \end{split}
    \label{eq:summation} \ .
\end{equation}
The results are
\begin{equation}
    \mu_{n^{*}}\sim -0.74\mu_N \ , \quad
    \mu_{p^{*}}\sim 1.55\mu_N \ ,
    \label{eq:estimate2}
\end{equation}
which are similar values with those
obtained in Ref.~\cite{Chiang:2002ah}
($\mu_{n^{*}}\sim -0.56\mu_N$
and $\mu_{p^{*}}\sim 1.86\mu_N$),
because in Eq.~\eqref{eq:summation}
the $K\Sigma$ component~($|g_{K\Sigma}|^{2}$)
dominates the $N(1535)$ resonance~($\sim$ 60 \%).
Naively, it is expected that this estimation corresponds to
the contribution from $-i\tilde{t}^{(b)}_{ij}$,
where the magnetic moments of the ground states are summed.

\subsection{Comparison with quark model}

Here we discuss the present results in comparison with the 
quark model results.  
The details how to compute the resonance magnetic moments 
have been presented previously~\cite{Chiang:2002ah}, and therefore, 
here we show some relevant points.  
In the quark model, the wave function of 
$N(1535)$ is given as a superposition of two 
spin ($s=1/2$ and 3/2) states in the $l=1$ 
70-dimensional representation of SU(6):
\begin{equation}
|N(1535)\rangle 
= \cos\theta |s=1/2\rangle + \sin \theta |s=3/2\rangle \, ,
\label{qmwf}
\end{equation}
where $\theta$ is a mixing angle of the twos states. 
Actually, the spin $s=1/2, 3/2$ states are coupled with the orbital 
angular momentum $l=1$ to yield $j=s+l=1/2$.  
The magnetic moment operator is a sum of spin and orbital 
angular momenta of three quarks, 
\begin{equation}
\mu = \frac{1}{2m} \sum_{i=1,2,3} \Bigl(
\sigma_{3}(i) + l_{3}(i)\Bigr) \, .
\end{equation}
By taking a matrix element between the quark model state 
(\ref{qmwf}), we obtain the magnetic moment as a function of 
the mixing angle $\theta$.

\begin{figure}[tbp]
    \includegraphics[width=6cm,clip]{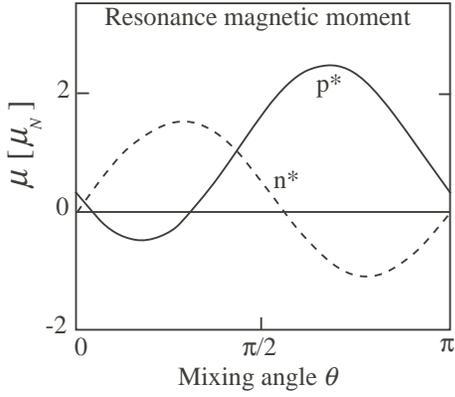}
    \caption{Magnetic moments as a function of 
    the mixing angle $\theta$
    in units of the nuclear magneton.}
    \label{fig:muquark}
\end{figure}

The result is presented in Fig.~\ref{fig:muquark}.
As reported in Ref.~\cite{Chiang:2002ah}, the mixing 
angle $\theta \sim 150 \sim -30$ degrees of 
the Isgur-Karl quark model yields the values
\begin{equation}
    \mu_{n^{*}} = -1.2 \mu_N \ , \quad
    \mu_{p^{*}} = 1.9\mu_N \ . 
\end{equation}
Although these numbers differ quantitatively from those in the chiral 
unitary model, they look similar qualitatively.  
In fact, it is interesting to observe that this happens 
only in the vicinity of 
the mixing angle $\theta \sim 150$ degree.
The similarity between the predictions in the 
quark model and the chiral unitary model was also reported for 
the axial coupling constant of $N^{*}$, or equivalently the 
$\pi N^{*}N^{*}$ coupling constant (due to the Goldberger-Treiman
relation)~\cite{Nacher:1998mi}.

\subsection{Magnetic moments in the chiral doublet model}

In this section, we present yet another description for
magnetic moments 
when the resonance of negative parity is regarded as 
a chiral partner of the ground state nucleon in linear 
representations of chiral symmetry.  
In addition to phenomenological aspects, 
such a point of view may shed a light on the properties of 
spontaneous breaking of chiral symmetry.  
The theoretical scheme for positive and negative parity nucleons 
has been discussed in detail in 
Ref.~\cite{Jido:1998av,Jido:2001nt}, and here we follow the essence of their description.   
The relevant point is that the chirality 
structure of the electromagnetic coupling; 
the vector coupling is of chirality even, 
while the tensor (anomalous magnetic)
coupling is of chirality odd:
\begin{align}
{\cal L}_{\gamma NN} =& 
\bar N \left(
\gamma_\mu + i \kappa_{\alpha} \frac{\sigma_{\mu \nu}q^\nu}{2\ProtonMass} \right) 
\tau_\alpha N A^\mu \nonumber \\
= & 
\left( \bar N_l \gamma_\mu \tau_\alpha N_l 
+
\bar N_r \gamma_\mu \tau_\alpha N_r \right) 
A^\mu\label{mugNN} \\
& %\hspace*{-2cm} 
+ \;  
i \kappa_{\alpha} \left( \bar N_l \frac{\sigma_{\mu \nu}q_\nu}{2\ProtonMass} 
\tau_\alpha N_r + \bar N_r \frac{\sigma_{\mu \nu}q_\nu}{2\ProtonMass} 
\tau_\alpha N_l \right)
 A^{\mu}\, .
\nonumber
\end{align}
Here $\alpha = 0$ or 3; 
$\tau_0 = 1$ is for the isoscalar and $\tau_3$ for the isovector 
components of the current.  The proton mass $\ProtonMass$ 
is just used for the unit.
The right and left handed components of the nucleon is defined by 
$
N_{r,l} = \frac{1 \pm \gamma_5}{2} N \, .
$

In the spirit of the theory of chiral symmetry, 
the electromagnetic coupling is regarded as 
a part of the chiral invariant coupling with 
right and left chiral fields.  
In Eq.~(\ref{mugNN}) the vector term preserves 
chiral symmetry, while the tensor (anomalous) term does not appear so. 
In order for the latter to be chirally symmetric, it should contain
the chiral field $U_5 = \sigma + i \vec \tau \cdot \vec \pi \gamma_5$.  
When chiral symmetry is broken spontaneously, $\sigma$ takes a 
finite expectation value $\bra \sigma \ket$, 
which survives the tensor term.  

Another interesting possibility is  
to construct a chiral invariant tensor term 
in the mirror model for positive and negative parity 
nucleons~\cite{Jido:1998av,Jido:2001nt}, where 
the basis of the chiral symmetry does not coincides to the physical
basis. Denoting the two chiral basis fields as $N_1$ and $N_2$,
the tensor coupling term takes on the form
\be
{\cal L}_{\gamma NN} = 
\frac{i \kappa}{2\ProtonMass}
\left(\bar N_1 \sigma_{\mu \nu} \gamma_5 N_2 
+ \bar N_2 \sigma_{\mu \nu} \gamma_5 N_1 \right) q^\nu A^\mu \, . 
\label{gNR_mirror}
\ee
This is the lagrangian to the lowest order ($n = 0$) 
in powers of 
$\bra \sigma \ket^n$ and is becoming a dominant term as
chiral symmetry is getting restored, 
$\bra \sigma \ket \to 0$.  Note again that the proton mass 
$\ProtonMass$ here is introduced only for the unit and has nothing
to do with the dynamical generated mass of nucleon in the linear sigma model.
In the following discussion, we consider only this 
leading order term of ${\cal O}(\bra \sigma \ket^0)$, 
in order to reduce the number of free parameters.  

As discussed in Ref.~\cite{Jido:1998av,Jido:2001nt}, 
the physical nucleon and $N(1535)$ fields 
are linear combinations of 
$N_1$ and $N_2$; 
$N(939) = \cos \theta N_1 + \gamma_5 \sin \theta N_2, 
N(1535) = - \gamma_5 \sin \theta N_1 + \cos \theta N_2$, 
where $\theta$ is a mixing angle.  
After diagonalization, the coupling term
takes on the form ($N_+ \equiv N(939), N_- \equiv N(1535)$):
\begin{align}
{\cal L}_{\gamma NN} =& 
\frac{i  }{2\ProtonMass} \left(
\sin 2\theta 
(\bar N_{+} \Gamma N_{+} + \bar N_{-} \Gamma N_{-} ) \right. \nonumber \\
&- \left. 
\cos 2\theta  (\bar N_+ \Gamma_5 N_- +  \bar N_- \Gamma_5 N_+)
\right) \, , 
\end{align}
where we have introduced the notation
$
\Gamma = \sigma_{\mu \nu} q^\nu 
(\kappa_S + \kappa_V \tau_3) A^\mu$, $
\Gamma_5 = \Gamma \gamma_5
$
with $\kappa_S$ and $\kappa_V$ being the isoscalar and isovector 
contributions to the anomalous magnetic moments.  
We find that the anomalous
%, and hence the total (the sum of the vector and tensor) 
magnetic moments 
of $N(939)$ and $N(1535)$ are the same; 
$\kappa_p = \kappa_{p^*}$, $\kappa_n = \kappa_{n^*}$.  
In the chiral unitary model, however, this is not 
the case.  

Let us now briefly discuss the transition moments.  
Note that the transition term has the structure of $E1$ because of 
the parity.  
>From the pion coupling strength of $N(1535)$ decay, 
the mixing angle was estimated as $\theta \sim 6.3$ 
degree~\cite{Jido:1998av,Jido:2001nt,DeTar:1989kn}.  
We can then use the diagonal components of the 
magnetic moments for the proton and neutron
to fix the $\kappa$'s:
$\kappa_S \sin 2\theta = -0.06$ and 
$\kappa_V \sin 2 \theta = 1.85$.  
Using these numbers, we find for the transition 
moments: 
$\mu_{pp^*} = 8.42$ and
$\mu_{nn^*} = -8.99$.  
The isovector dominance in these quantities is consistent with
what is known from experiment, but the magnitudes of these
numbers are too large as compared with experimental data, 
$|\mu_{pp^*}| \sim 2 |\mu_{nn^*}| \sim 1$
in units of the nuclear magneton, as extracted from 
the helicity amplitudes
$A_{1/2}^{p,n} \sim 90, -46$ $10^{-3} {\rm GeV}^{-1/2}$~\cite{Hagiwara:2002fs}.  

Phenomenologically, both diagonal and 
transition moments do not agree with data.  
In particular, the small mixing angle has lead to the 
dominance of the off-diagonal components, which should be 
so in the world where chiral symmetry breaking is not 
so strong.   
We can make some speculations about the 
nature of the breaking of chiral symmetry.  
For instance, the fact that $\mu_N \neq \mu_{N^*}$ 
could be an indication that higher order terms in $\bra \sigma \ket$ 
should be important.  
The large transition moments may suggest
a larger mixing angle, as opposed to the result 
obtained from the pion couplings
previously~\cite{Jido:1998av,Jido:2001nt,DeTar:1989kn}.  
Both facts may be an indication that chiral symmetry is broken 
rather strongly.  
In any event, magnetic moments of the nucleon as well 
as of its excited state provide useful information 
of chiral symmetry of baryons.

%%%%%%%%%%%%%%%%%%%%%%%%%%%%%%%%%%%%%%%%%%%%%%%%%%%%%%%%%%%
\section{Observation of the $N^{*}$ magnetic moment}
\label{sec:exp}

%%%%%%%%%%%%%%%% FIGURE %%%%%%%%%%%%%%
\begin{figure*}[tbp]
\epsfxsize =15cm
\epsfbox{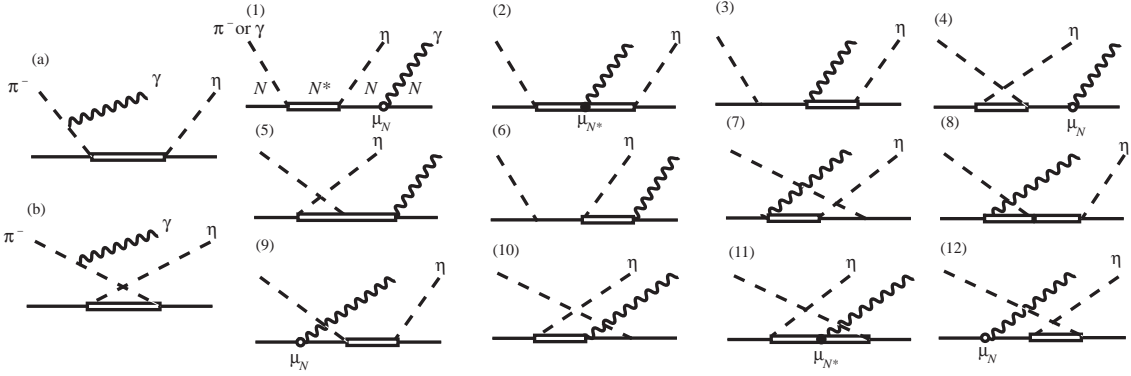}
\caption{The relevant diagrams for the processes $\gamma N \rightarrow 
\gamma \eta N $ and $\pi^{-} p \rightarrow \gamma \eta n$. The diagram 0
is only for the pion induced process. 
The initial boson is ether photon or pion. 
The solid and double-solid lines denote $N$ and $N^{*}$, respectively.
The dotted and wavy lines in the final state are
the emitted photon and eta meson, respectively. 
\label{fig:12dias}}
\end{figure*}
%%%%%%%%%%%%%%%% FIGURE %%%%%%%%%%%%%%

In this section, we would like to discuss possibilities of the experimental 
observations of the $N^{*}$ magnetic moments. The $N(1535)$ has the
special feature that this resonance strongly couples to the $\eta N$ system,
which is not seen in the other $N^{*}$ resonances.
Thus the $\eta$ meson in the finial 
state may be regarded as a probe of $N(1535)$ in the intermediate state.
In order to observe the $N(1535)$ magnetic moments,
here we would like to calculate
cross sections of the following two photon-emission processes; 
$\gamma N \rightarrow \gamma \eta N$ and 
$\pi^- p \rightarrow \gamma \eta n$, and we investigate sensitivity
of their cross sections to the value of the magnetic moments of $N^{*}$.
Such processes that two-boson emission on
nucleon target are discussed in Ref.~\cite{Jido:2001nt,Jido:2000nt} to observe the 
sign of the $\pi N^{*}N^{*}$ coupling. In the present work, 
we follow their method to calculate the cross sections of the above processes.

In the calculations of the cross sections,
we use the Lagrangian formulation for $N^{*}$, where the $N^{*}$ is 
described as a well-defined field and its propagator is assumed to be 
the Breit-Wigner form with the mass $\NstarMass = 1535$ MeV and the
width $\NstarWidth = 150$ MeV.
The Lagrangians  used in the present calculations are shown 
in Appendix~\ref{app:Lag}.

Now we assume the $N^{*}$ dominance hypothesis in the $\eta$-$N$ system near
the threshold region, that is, 
the $\eta$ meson can couple only to the $N$ and $N^{*}$ transition, and
the other resonances do not couple to the $\eta$ meson. 
It is shown in Ref.~\cite{Jido:2001nt} 
that this hypothesis reproduce the $N(1535)$ resonance 
well in the $\pi N \rightarrow \eta N$ process. 
Then the relevant
diagrams for these processes are shown in the Fig.~\ref{fig:12dias}. The diagrams
$a$ and $b$ are used only for the pion-induced process. 
Since we consider the photon-eta production processes in the energies close to 
the threshold, the final photon and eta meson have small energies and, therefore, 
the dominant contributions come from the diagrams $a$, 1, 2, 9 
as a result of their small energy denominators. 
The diagram 2
is the one in which the magnetic moment of $N^{*}$ appears, and we want 
to see the interference effects of this diagram with the other dominant diagrams
to study the $N^{*}$ magnetic moments.

The differential cross section is given as
\begin{equation}
  d \sigma = \frac{2 \Nmass}{4 \sqrt{(p_{i}\cdot k_{i})^{2}-
   \Nmass^{2} m_{IB}^{2}}} \frac{1 }{2}\frac{1}{ 2} \sum_{\stackrel{\rm spin}
   {\rm pol.}} |{\cal T}_{fi}|^{2} d\Phi \ ,
\end{equation}
where the summation is taken over the spin of the initial and final nucleons and
the polarization of the final (and initial) photon, and the factors in front of the 
summation are for taking averages of the spin and polarization in the final state.
The mass $m_{IB}$ denotes the mass of the initial boson, photon or pion in the
present case.
The phase space of the three-particle state is given by
\begin{equation}
\begin{split}
 d\Phi = & (2\pi)^{4} 
  \delta(p_{i} + k_{i} - p_{f} - k_{\gamma} - k_{\eta}) \\
  &\times \frac{d^{3} \vec k_{\gamma}}{ (2\pi)^{3}2E_{\pi}}
   \frac{d^{3}  \vec k_{\eta}}{(2\pi)^{3}2E_{\eta}}
    \frac{\Nmass d^{3} \vec p_{f} }{ (2\pi)^{3}2E_{f}} \ , \label{eq:phase}
 \end{split}
\end{equation}
where $k_{\gamma}=(E_{\gamma},\vec k_{\gamma})$, $k_{\eta}=(E_{\eta}, \vec k_{\eta})$ and $p_{f}=(E_{f}, \vec p_{f})$ are momenta for the finial photon,
eta and nucleon, respectively.
In the center of mass frame, Eq.~(\ref{eq:phase}) is written as
\begin{equation}
  d\Phi = \frac{\Nmass}{4 (2\pi)^{5}}  dE_{\gamma} dE_{f} d\alpha d(\cos \beta)
  d\gamma \ .
\end{equation}
Here $\alpha, \beta, \gamma$ are the Eular angles, which specify the plane where
the three momenta in the final state lie.
The normalizations of the state and wave function for nucleon are 
\begin{equation}
\bar u^{(\alpha)}(p) u^{(\beta)}(p) = \delta^{\alpha\beta} \ ,
\end{equation}
\begin{equation}
\langle p | p^{\prime} \rangle = \frac{E}{ \Nmass} (2\pi)^{3} 
\delta^{3}(\vec p - \vec p^{\prime}) \ .
\end{equation}

In the calculations of the cross sections, we perform the integral over
the three-body phase space with the Monte Carlo method. The number of 
the integration points in the present calculations are taken larger than 10,000, which
may be enough to converge the Monte Carlo integral. The details of the Monte
Carlo method for the three-body final state are discussed in Ref.~\cite{Jido:2001nt}.

\begin{figure}[tbp] 
\epsfxsize=8cm
\epsfbox{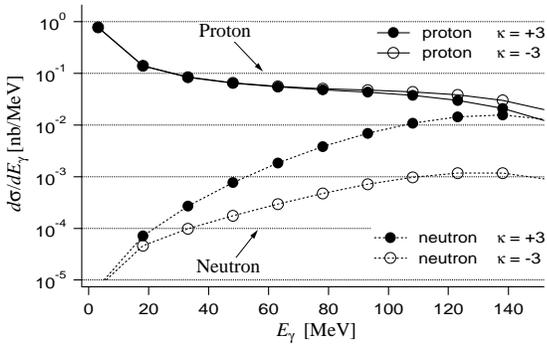}
\caption{Energy spectra of the emitted photon in the photon-induced process.
The energy of the initial photon is 1000 MeV. The solid and dotted lines denote
the proton and neutron targets, respectively. The anomalous magnetic moments 
of $p^{*}$ and $n^{*}$ are assumed to $+3$ (black circle) or $-3$ (white circle)
in units of nuclear magneton.
\label{fig:photoSpec}}
\end{figure}

Shown in Fig.~\ref{fig:photoSpec} are the energy spectra of the emitted photon
in the photon-induced processes with the proton and neutron targets;
$\gamma N\to \gamma\eta N$. 
In order to see the sensitivity of the effect of the magnetic 
couplings of $N^{*}$ to the cross sections, the 
anomalous magnetic moments for the nucleon resonances are assumed to be 
$+3$ or $-3$ in units of nuclear magneton, although the predicted values by the
present work are much smaller.

%%%%%%%%%%%%%%%% FIGURE %%%%%%%%%%%%%%
%%%%%%%%%%%%%%%% FIGURE %%%%%%%%%%%%%%

\begin{figure}[tbp]
\epsfxsize=8cm
\epsfbox{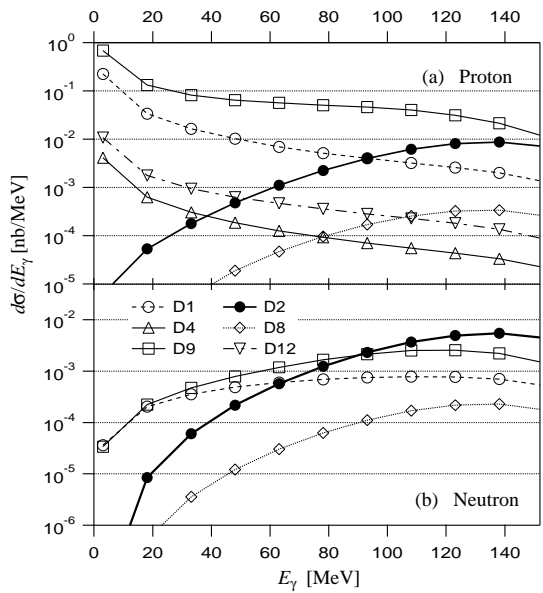}
\caption{Separated contributions for the dominant diagrams to the energy spectra in 
the photon-induced process with the initial photon energy $E_{i}=1000$ MeV. 
The upper panel (a) is for the proton target case. The lower panel (b) is the same
for the neutron target case. The anomalous magnetic moments of 
$p^{*}$ and $n^{*}$ are $+3 \mu_{N}$. The lines with black circles, 
open circles, triangles, diamonds, 
squares and down-triangles denote the contributions from the diagrams
1, 2, 4, 8, 9 and 12, respectively.  
\label{fig:photoSpecCont}}
\end{figure}

In the case of the proton target,
where we investigate the magnetic moment of $p^{*}$, the  resonance
magnetic moment is not sensitive to the energy spectrum of the emitted photon,
as shown in Fig.~\ref{fig:photoSpec}.
Presenting the separated contribution from the diagram 2, 
in which the $N^{*}$ magnetic moment 
is involved % (*** or another word ***)
, we show in Fig.~\ref{fig:photoSpecCont} 
the energy spectra calculated with each dominant diagram of Fig.~\ref{fig:12dias}.
In Fig.~\ref{fig:photoSpecCont}, it is seen that the energy spectrum for the proton
target case is dominated by the contribution from the diagram 9 in all energies, which
corresponds to the bremsstrahlung of the initial proton. In the bremsstrahlung, the 
cross section becomes larger with the faster charged particle and 
the softer emitted photon.
The diagram 2 gives much smaller contribution than the diagram 9.
Therefore the 
$\gamma p\rightarrow \gamma\eta p$ is not appropriate process to observe
the $N^{*}$ magnetic moment. In fact, in the case of the charged particle proton, 
the electric coupling of the proton to the photon 
gives larger contributions than the magnetic one, since the magnetic coupling
linearly depends on the photon momentum and, hence,  is suppressed 
in energies near the threshold. 

%%%%%%%%%%%%%%%% FIGURE %%%%%%%%%%%%%%
%%%%%%%%%%%%%%%% FIGURE %%%%%%%%%%%%%%

On the other hand, in the case of the neutron
target, the sensitivity of the magnetic moment is seen in the higher energies
of the spectra as a result of the interference effects. 
Here we would have chance to observe the magnetic moment of 
$n^{*}$, although the cross sections are quite small and the all participants
in this reaction are neutral particles. As shown in Fig.~\ref{fig:photoSpecCont},
the diagram 9 is less dominant than the proton target case,
since there are no electric couplings for the neutron target case.
Rather than the amount of the cross section, however, distinct signals of
the dependence of the magnetic moments of $N^*$, such as position
of peak, are not seen in the energy spectra of the emitted photon.

\begin{figure}[tbp]
\epsfxsize=8cm
\epsfbox{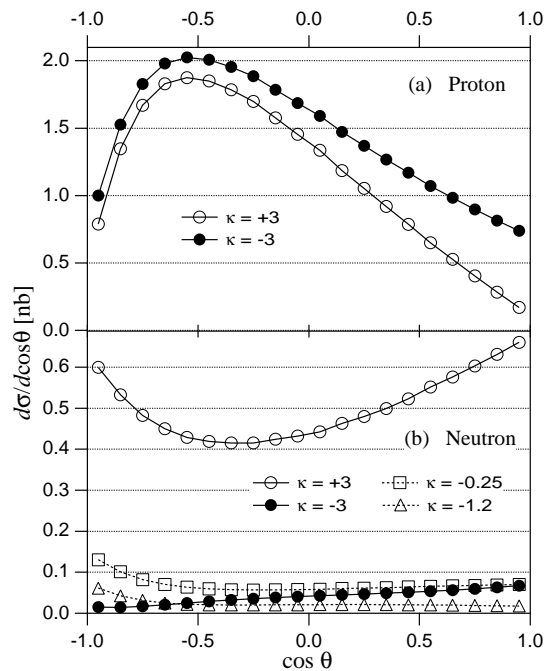}
\caption{Angular distributions of the emitted photon in the photo-induced
process with the initial photon energy 1000 MeV. 
The integration with respect to the
emitted energy is performed from 80 MeV to the threshold. 
The upper panel (a) is for the proton target, and the lower panel (b) is the same for
the neutron target. 
The anomalous magnetic moments of $N^{*}$
is assumed to $+3$ (open circle) or $-3$ (black circle) in units of nuclear magneton.
In the case of the neutron target, the plots with
the $N^{*}$ magnetic moments obtained in the chiral unitary model ($\mu^{(a)}=-0.25$) 
and in the quark model ($\mu^{(a)}=-1.2$) cases are shown by
the lines with open squares and triangles. respectively.
\label{fig:angPhoto}}
\end{figure}

\begin{figure}[tbp]
\epsfxsize=8cm
\epsfbox{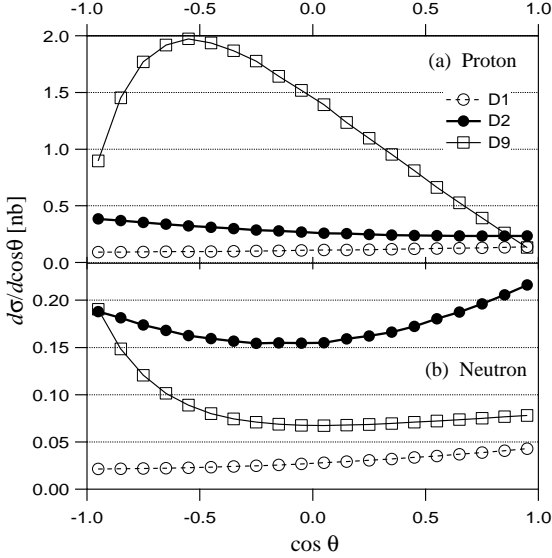}
\caption{Separated contributions of the dominant diagrams to the angular distribution
of the emitted photon in the photon-induced process. 
The integration with respect to the
emitted energy is performed from 80 MeV to the threshold. 
The lines with the open circles, black circles and squares denotes the contributions
of the diagrams 1, 2, and 9, respectively. 
\label{fig:angPhotoCont}}
\end{figure}

Next we calculate the angular distributions of the emitted photon, which
is expected to be a better example to see the interference effects.
Shown in Fig.~\ref{fig:angPhoto} are the calculated angular distributions in terms
of the angle $\theta$ between the incident and final photons.
Here we find the distinct angular
dependences in the case of the neutron target, which would be observed.
We also plot in Fig.~\ref{fig:angPhoto}(b) the angular distributions with
the $N^{*}$ magnetic moments obtained by the chiral unitary model and the
quark model. It might be difficult, however, to distinguish these two model 
in experiment.  
In the calculations of the angular distributions, we perform the integration 
with respect to the final photon energy from 80 MeV, since we want to 
see the interference effects of the diagram 2 to the others and the diagram~2
gives dominant contributions at photon energies larger than 80 MeV as seen in
Fig.~\ref{fig:photoSpecCont}.
We show in Fig.~\ref{fig:angPhotoCont} the separated contributions to
the angular distributions to the emitted photon.
As seen in the figure, the diagram 2 becomes the dominant diagram in the
case of the neutron target, while, 
in the proton target case,  the diagram 9 is still the most dominant diagram.

\begin{figure}[tbp]
\epsfxsize=8cm
\epsfbox{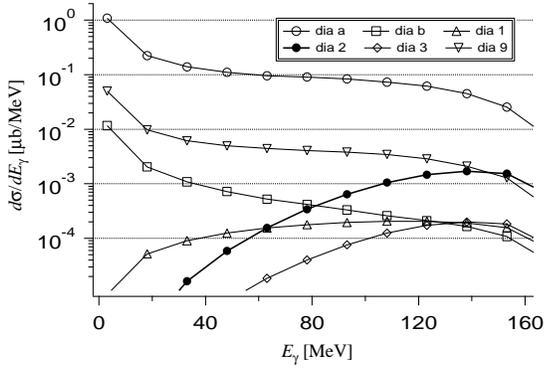}
\caption{Separated contributions of the dominant diagrams to the energy spectrum of 
the emitted photon in the pion-induced process with the initial pion energy
$E_{\pi}=1000$ MeV. 
The lines with the open circles, squares, triangles, black circles, diamonds and
down-triangles denote the contributions from the diagram $a$, $b$, 1, 2, 3 and 9,
respectively. 
\label{fig:pionInduced}}
\end{figure}

Finally we discuss the pion-induced process briefly. 
As discussed before, in the case of neutron, the value of the $n^{*}$ magnetic moment
is sensitive to the cross sections, since the magnetic contributions is relatively 
enhanced due to the absence of the electric coupling. Thus, we would expect that 
the $\pi^{-}p \rightarrow \gamma\eta n$ process would be good to observe
the magnetic moment of $n^{*}$. Unlike our expectation, however, 
in this case, we conclude that it is very difficult to extract the magnetic moments of
$N^{*}$, since the diagrams $a$ and 9 are the most dominant contributions to the
cross sections, as shown in Fig.~\ref{fig:pionInduced}. These diagrams corresponds
to the bremsstrahlung of the initial charged particles. Since the initial pion and proton
have large momenta to create the eta meson at the final state, 
they emit the more photon than the slow intermediate $n^{*}$.  

%%%%%%%%%%%%%%%%%%%%%%%%%%%%%%%%%%%%%%%%%%%%%%%%%%%%%%%%%%%
\section{Summary}\label{sec:summary}

We have calculated the magnetic moments of the $N(1535)$ resonance
using the chiral unitary model.
We have obtained the magnetic moments of the resonances as
$\mu_{n^{*}(1535)}\sim -0.25\mu_N$ and
$\mu_{p^{*}(1535)}\sim +1.1\mu_{N}$.
Compared with the results of $\Lambda$ resonances
in Ref.~\cite{Jido:2002yz},
the sign of the Coleman-Glashow relations~\eqref{eq:CGrelations},
which comes from the SU(3) symmetry of octet,
are satisfied among $\Lambda^*(1670)$ and $n^{*}(1535)$
in the chiral unitary model.
This implies that $1/2^{-}$ resonances are the member
of an SU(3) octet.
The present results qualitatively
agree with the results of the constituent quark model
of Ref.~\cite{Chiang:2002ah}. 
However, the absolute values of these results are different,
so the experimental measurement
will bring the information of the structure of the baryon resonances.
Finally we have computed reaction cross section 
in order to observe the resonance magnetic moments;
$\gamma N\to \gamma \eta N $, $\pi^-p\to \gamma\eta n$.
The difference in the magnetic moments is,
however, not very much reflected in the bremsstrahlung processes.

\begin{acknowledgments}
We would like to thank Prof.~E.~Oset,
for checking the calculations in detail.
We would like to acknowledge Profs.~M.~J. Vicente~Vacas
and H.\ -Ch.\ Kim for useful discussions.
\end{acknowledgments}

\appendix
\section{Lagrangians}
\def\theequation{\Alph{section}.\arabic{equation}}
\setcounter{equation}{0}
\label{app:Lag}

In this appendix, we  show the Lagrangians used in Sec.~\ref{sec:exp}
to calculate the cross sections of the $\gamma N \rightarrow \gamma \eta N$
and $\pi N \rightarrow \gamma \eta N$ processes though the $N^{*}$ intermediate 
state. Here $N^{*}$ denotes the $N(1535)$ resonance, which has the negative parity.

For the $\eta NN^{*}$ vertex, we take the scalar coupling:
\begin{equation}
{\cal L}_{\eta NN^{*}} = g_{\eta} \bar{N} N^{*} + {\ \rm h.c.} \ ,
\end{equation}
where the coupling constant $g_{\eta}\simeq 2.0$ is determined so as to reproduce
the partial decay width $\Gamma_{N^{*}\rightarrow \eta N} \simeq 75$ 
MeV~\cite{Hagiwara:2002fs} at tree level.

The transition vertex of $N$ to $N^{*}$ with one photon is given by
\begin{equation}
 {\cal L}_{\gamma NN^{*}} = \frac{ e }{ 4 \ProtonMass} 
\mu^{(T)}_{N} \bar N i\gamma_{5} \sigma^{\mu\nu} F_{\mu\nu} N + {\ \rm h.c.}\ . 
\end{equation}
Here we assume the isovector dominance on the transition magnetic 
moments $\mu^{(T)}_{N}$, and their values are given by $\mu^{(T)}_{p} 
= - \mu^{(T)}_{n} = 0.68$ in units of nuclear magneton, which correspond 
to $\kappa^{*}_{V}\equiv \frac{2\Nmass}{\Nmass+\NstarMass} \mu^{(T)}_{N} 
= 0.9$ determined from analyses of eta
photoproduction~\cite{Benmerrouche:1995uc}.

The $\gamma NN $ and $\gamma N^{*}N^{*}$ vertices have two parts, which are
so-called the Dirac term and the Pauli term:
\begin{align}
    &
    \begin{split}
    {\cal L}_{\gamma NN} =& -eQ \bar N \gamma_{\mu} A^{\mu} N \\
       &+ \frac{e}{4 \ProtonMass} \kappa_{N}
\bar N \sigma^{\mu\nu}F_{\mu\nu} N \ , 
    \end{split} \\
    &
    \begin{split}
    {\cal L}_{\gamma N^{*}N^{*}} = &- e Q\bar N^{*} \gamma_{\mu} A^{\mu} N^{*} \\
    + &\frac{e }{ 4 \ProtonMass}\kappa_{N^{*}} \bar N^{*} \sigma^{\mu\nu}
    F_{\mu\nu} N^{*} \ .
    \end{split}
\end{align}
The anomalous magnetic moments of 
the ground state nucleons $\kappa_{N}$ 
are used the experimental value $\kappa_{p}=1.79284739$ and 
$\kappa_{n}=-1.9130428$ in units of nuclear magneton~\cite{Hagiwara:2002fs}, while
the anomalous magnetic moments of $N^{*}$ are assumed to be $\pm 3$ to see
sensitivity of their values to the cross sections.

For the calculations of the pion-induced process, we use the following Lagrangians. 
The $\pi NN^{*}$ vertex has the scalar type coupling, which given by
\begin{equation}
{\cal L}_{\pi NN^{*}} =g_{\pi NN^{*}} \bar N \vectau \cdot \vecpi N^{*} 
+ {\ \rm h.c.} \ .
\end{equation}
with the coupling constant $g_{\pi NN^{*}}\simeq 0.7$, which is determined so as
to reproduce the partial decay width $\Gamma_{N^{*}\rightarrow \pi N} \simeq
75$ MeV~\cite{Hagiwara:2002fs} at tree level. The diagonal vertices $\pi NN$ 
and $\pi N^{*}N^{*}$ are assumed to be here the pseudo-scalar couplings:
\begin{eqnarray}
{\cal L}_{\pi NN} &=& g_{\pi NN} \bar N \gamma_{5} \vectau 
\cdot \vecpi N \ , \\
{\cal L}_{\pi N^{*}N^{*}} &=& g_{\pi N^{*}N^{*}} \bar N^{*} \gamma_{5}
\vectau \cdot \vecpi N^{*}\ , 
\end{eqnarray}
Here we use the empirical value of the $\pi NN$ coupling $g_{\pi NN} \simeq 13$.
For the $\pi N^{*}N^{*}$ coupling, we assume $g_{\pi N^{*}N^{*}} \simeq +13$, 
although the sign of this coupling is important for the properties of $N^{*}$. 
The value of the $\pi N^{*}N^{*}$ coupling is absolutely insensitive to the final
results, since this coupling appears in the less dominant diagrams.
In this formulation, we do not include the Kroll-Ruderman type diagram, since 
we use the scalar type coupling for the $\pi NN^{*}$ vertex and it has already
contain the partial contribution of the Kroll-Ruderman type vertex.   
The pion-photon coupling is given by
\begin{equation}
\begin{split}
{\cal L}_{\gamma\pi\pi} =  &i e (\del_{\mu} \pi^{-})^{\dagger} \pi^{-} 
A^{\mu}  \\
& - i e \pi^{-\dagger}(\del_{\mu} \pi^{-}) A^{\mu}
\end{split}
\ .
\end{equation}

\end{document}